\documentclass[preprint,12pt]{elsarticle}
\usepackage{amssymb}
\usepackage{lineno}
\usepackage{amsmath}

\journal{Nucl. Instr. and Meth. Phy. Res. Sect. A}

\begin{document}
\begin{frontmatter}

\title{Gamma Ray Detection Efficiency of GAGG Crystal Scintillator Using Three Tagged Gamma Ray Techniques}

\author[inst1]{B. Doty\fnref{bDoty}\corref{cor1}}
\fntext[bDoty]{\textit{Email:} bsdoty@shockers.wichita.edu}

\author[inst1]{N. Solomey}

\author[inst1]{J. Folkerts}

\author[inst1]{B. Hartsock}

\author[inst1]{H. Meyer}

\author[inst2]{M. Christl}
\author[inst2]{M. Rodriguez-Otero}
\author[inst2]{E. Kuznetsov}

\affiliation[inst1]{organization={Wichita State University},
            addressline={1845 Fairmount St.}, 
            city={Wichita},
            postcode={67260}, 
            state={Kansas},
            country={United States of America}}

\affiliation[inst2]{organization={Marshall Space Flight Center, NASA},
            addressline={320 Sparkman Blvd.}, 
            city={Huntsville},
            postcode={35812}, 
            state={Alabama},
            country={United States of America}}

\begin{abstract}
A CubeSat with a prototype scintillating detector with a sensitive volume of Gadolinium-Aluminum-Gallium-Garnet crystal is being developed with a possible launch date of 2025. Its purpose is to characterize the background signals that mimic the neutrino interaction that the $\nu$SOL (Neutrino Solar Orbiting Laboratory) team is looking for. An important part of the characterization of the backgrounds and the expected real signal is understanding the gamma ray photopeak efficiency of the prototype detector when compared to simulations performed in Geant4. To this end we have used three techniques to do a measurement of the  gamma ray efficiency compared to simulation. The first is using electron capture sources that emit an X-ray before prompt emission of a de-excitation gamma ray, specifically $^{65}$Zn and $^{54}$Mn. The second is using a $\beta^+$ decay source wherein a positron annihilates on an atomic shell electron producing two back-to-back 511 keV gammas followed promptly by a de-excitation gamma, specifically $^{22}$Na. The third is using a gamma cascade of two near-simultaneous de-excitation gammas from the same nucleus, specifically from $^{60}$Co decay.
\end{abstract}

\begin{highlights}
\item We can determine a scintillator's photopeak efficiency compared to Geant4 simulations using radiation cascades.
\item GAGG crystal scintillators are a promising target for a neutrino detector. 
\item GAGG scintillation detectors can be up to 92.8\% efficient for gamma  detection when compared to Geant4 simulations.
\end{highlights}

\begin{keyword}
neutrino \sep gallium \sep GAGG \sep gamma \sep scintillating crystal \sep detector efficiency
\end{keyword}

\end{frontmatter}


\section{Introduction}\label{sec:Introduction}
The $\nu$SOL (Neutrino Solar Orbiting Laboratory) project aims to detect solar neutrinos with a space-based detector featuring a Gadolinium-Aluminum-Gallium-Garnet (GAGG) scintillating crystal using the double pulse delayed coincidence of the $^{71}$Ga neutrino absorption process \cite{SOLOMEY2023168064}. When a solar neutrino interacts with $^{71}$Ga, it promotes the nucleus to $^{71}$Ge, which may be excited so long as the neutrino has sufficient energy to produce the excited state. This excited state decays via a 175 keV gamma ray \cite{ref:BahcallGallium}, it is this gamma ray and the prompt electron that we aim to use as the neutrino identification technique. It is therefore critical that we understand the gamma ray photopeak detection efficiency of the GAGG detector when compared to simulation so that simulations can be used to analyze the eventual neutrino signals. It is important to mention that the 175 keV gammas almost exclusively produce full energy deposits in the flight detector and so we are interested in the measured vs. simulated photopeak efficiency and are not interested in the Compton scattering region. The desired energy value of 175 keV did not have any nearby gamma energies to use as an efficiency measurement from our lab apart from $^{57}$Co, which has a 136 keV and 122 keV gamma triggered by electron capture with an X-ray similarly to $^{65}$Zn and $^{54}$Mn. However there were unexpected theoretical complications with using this source to perform the efficiency measurement, namely a very strong angular correlation.

\begin{figure}[htbp]
\centering
\includegraphics[width = 0.6\textwidth]{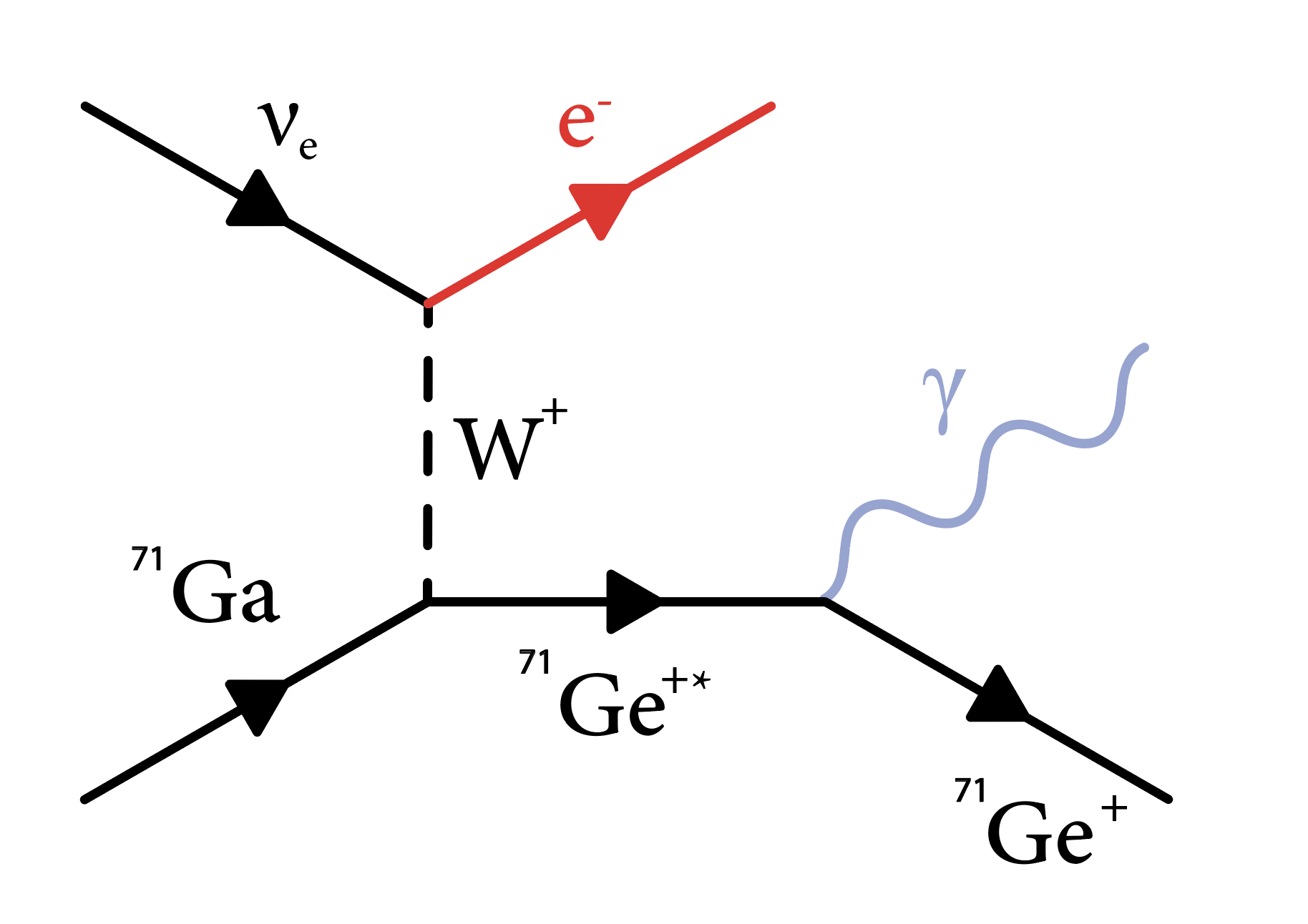}
\caption[Ga71 Interaction]{The relevant neutrino interaction to detect solar neutrinos. The $^{71}Ga$ nucleus interacts with the electron neutrino to produce an excited state of $^{71}Ge$, which then subsequently decays via a gamma ray.}
\label{fig:Ga71Interaction} 
\end{figure}

The overall mission has 2 stages. The second stage is a solar probe with many kilograms of GAGG being launched into a highly elliptical orbit with a closest approach of approximately 7 solar radii, following in the footsteps of Parker Solar Probe. It is this stage that will actually have a chance at detecting solar neutrinos. The first stage, in contrast to a large solar probe, is a 3U CubeSat with about 73 grams of GAGG that is launched into a polar orbit where the earth's magnetic shielding effect is small and so the detector will be exposed to background events that are expected to be similar to the backgrounds the eventual solar probe will experience. This detector will not detect a solar neutrino, but will give a real look into the backgrounds that could potentially interfere with the neutrino reconstruction process. The gamma detection efficiency versus simulation is necessary for characterizing said backgrounds.

We have determined the gamma detection efficiency compared to simulation for the CubeSat flight detector, which is is a 2x2 array of GAGG crystals each 14x14x14 mm with a single quartz light guide that is 28x28x10 mm and 16 silicon photo-multipliers (SiPMs) on the quartz. There is optical resin between each of the GAGG and quartz surfaces.

\begin{figure}[htbp]
\centering
\includegraphics[width = 0.6\textwidth]{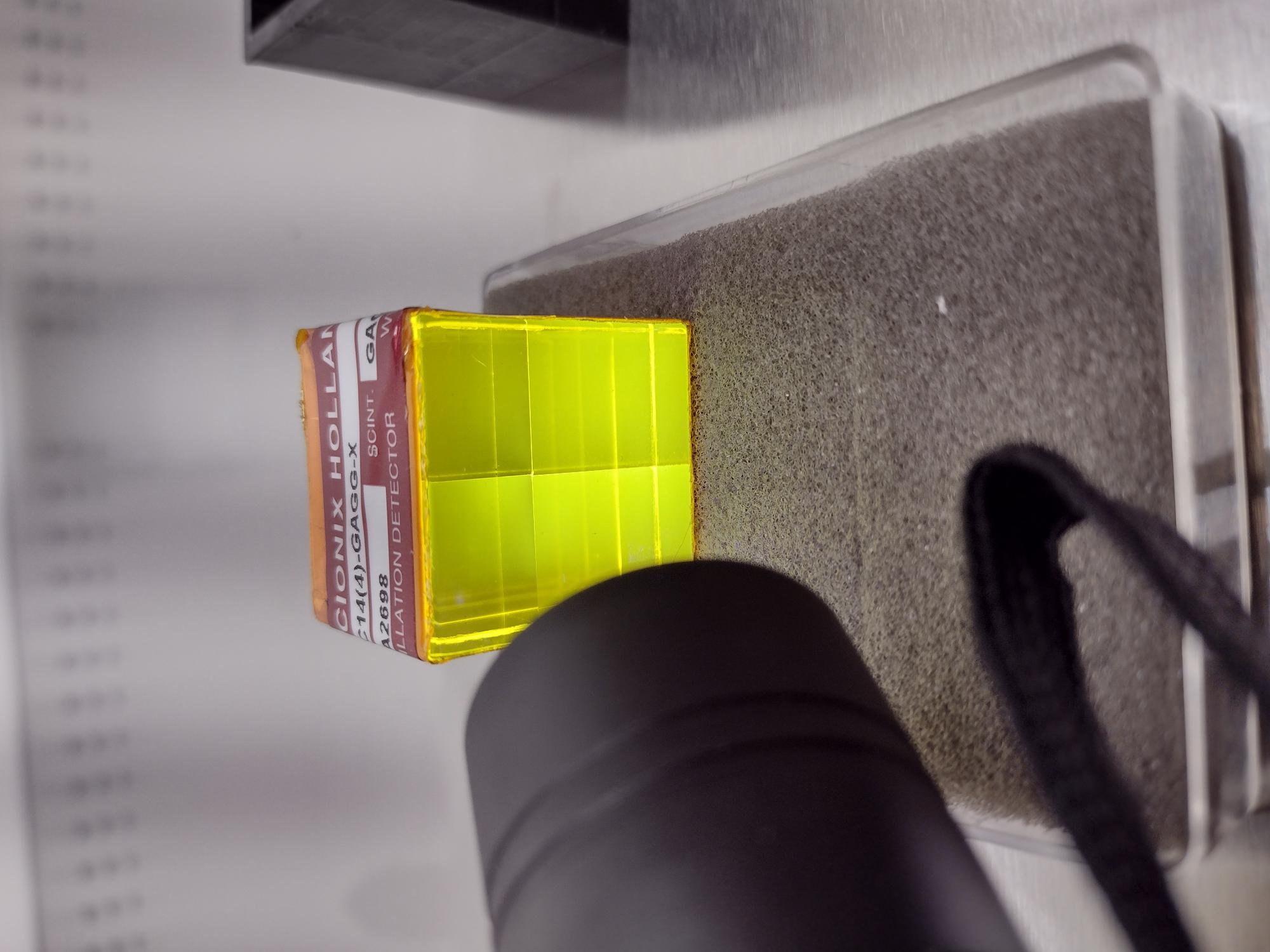}
\caption[GAGG Crystal Flight Detector]{The array of GAGG crystals under UV light. This is before the bonding with SiPMs and the side facing the viewer is the quartz light guide.}
\label{fig:GAGGUV} 
\end{figure}

\section{Experimental Technique and Equipment}\label{sec:ExperimentalTech}

\subsection{Triggered Gamma Ray Sources}\label{sec:X-rayTrigger}
\begin{figure}[htbp]
\centering
\includegraphics[width = 0.9\textwidth]{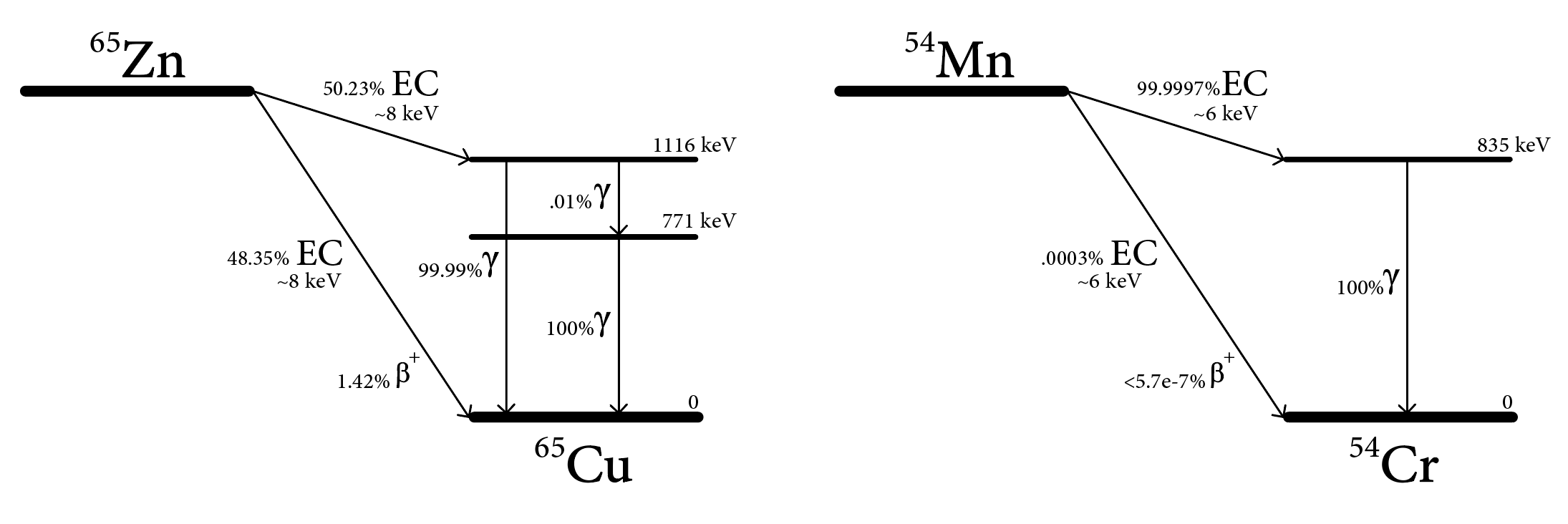}
\caption[$^{65}$Zn and $^{54}$Mn decay sequence.]{On the left is the $^{65}$Zn electron capture decay chain. On the right is the $^{54}$Mn electron capture decay chain.}
\label{fig:Zn and Mn Decay Schemes} 
\end{figure}

To determine the measured vs simulated gamma detection efficiency of a prototype detector, we must first know whether or not a gamma was even available to be detected. To do this, we use a triggered gamma ray source. This could be an X-ray triggered gamma source, an annihilation-triggered gamma source, or a gamma-triggered gamma source. We use an X-ray detector to trigger on one of the X-rays or gammas and look for the corresponding gamma in the flight detector, opposite to a similar absolute X-ray detection efficiency study that used the gamma ray as a trigger to look for X-rays in a transition radiation detector \cite{GRAHAM1995224}. For X-ray triggered gamma sources, both $^{65}$Zn \cite{A65} and $^{54}$Mn \cite{A54} undergo an electron capture (EC) of one of the atomic orbital electrons, which results in the emission of an X-ray, see Figure \ref{fig:Zn and Mn Decay Schemes}. After electron capture, the resultant nucleus is in an excited state with a known probability, it is this gamma from the excited nucleus that we look for in the prototype detector.

We can also use gamma-triggered gamma sources where the trigger is made from a gamma ray, either by an annihilation gamma or a gamma cascade; $^{22}$Na \cite{A22} and $^{60}$Co \cite{A60} are both viable for this technique as they both have multiple coincident gammas that can be used for an efficiency measurement. In the case of $^{22}$Na we used one of the 511 keV annihilation gammas to look for the 1274 keV gamma and the case where both the 1274 keV and 511 keV interact, which will hereafter be referred to as the combined peak. In the case of $^{60}$Co we can use either gamma ray as a trigger to look for the other gamma, though some care needs to be taken to avoid the Compton scattering backgrounds since the two gammas are so close in energy.

\begin{figure}[htbp]
\centering
\includegraphics[width = 0.9\textwidth]{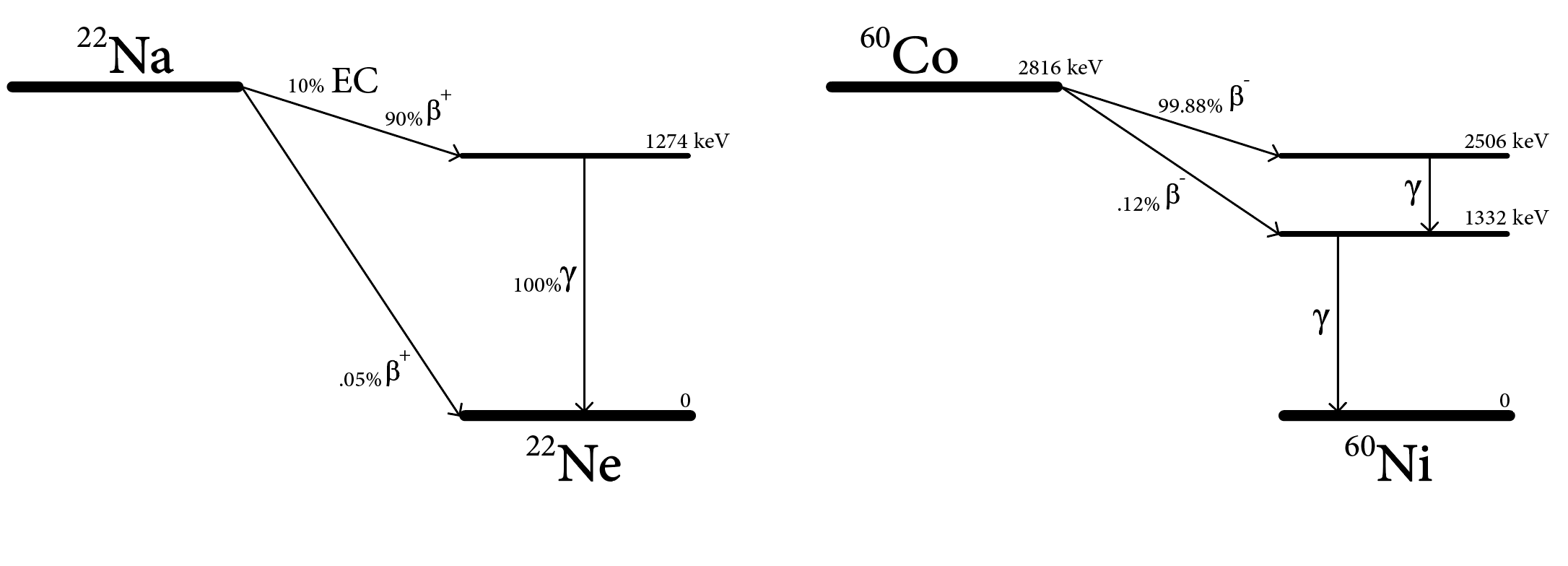}
\caption[$^{22}$Na and $^{60}$Co decay sequence.]{On the left is the $^{22}$Na decay chain. On the right is the $^{60}$Co decay chain.}
\label{fig:Na and Co Decay Schemes} 
\end{figure}

\subsection{Trigger Detector and Flight Detector Details}\label{sec:TriggerDetector}
The X-ray detector, 51$\phi$x10 mm$^3$, that is used as a trigger for this experiment is a Scionix detector that features a CeBr$_3$ crystal with a high light yield and a 0.3mm thick beryllium entrance window having a 250 ps FWHM resolving time. The light is readout using a PMT internal to the detector. This composition makes it suitable for detecting X-rays down to 5 keV and gamma rays up to about 2 MeV (containment limited by the 10 mm thickness). There is signal shaping and amplification internal to the detector.

The flight detector features 4 total crystals readout by 16 SiPMs. The signals are summed and shaped using a standalone electronics board developed by NASA Marshall Space Flight Center. A feature of this board is signal integration as the final output signal (this is why the signals from the flight detector are so much wider than the signals from the X-ray detector). The methods of signal shaping do not modify any of the results of this paper since the coincidence between the two detectors is determined within a time window that fully contains both pulses.

\subsection{Experimental Setup \& Acquisition Methods}\label{sec:ExperimentalSetup}
For each detector and for each run the following overall setup is used:

First, the source is taped directly to the X-ray detector, making sure that the side that is permeable to X-rays is facing the beryllium entrance window. Then, the combined source and X-ray detector is then brought as close as possible to the prototype detector in a manner that is reproducible.

\begin{figure}[htbp]
\centering
\includegraphics[width = 0.6\textwidth]{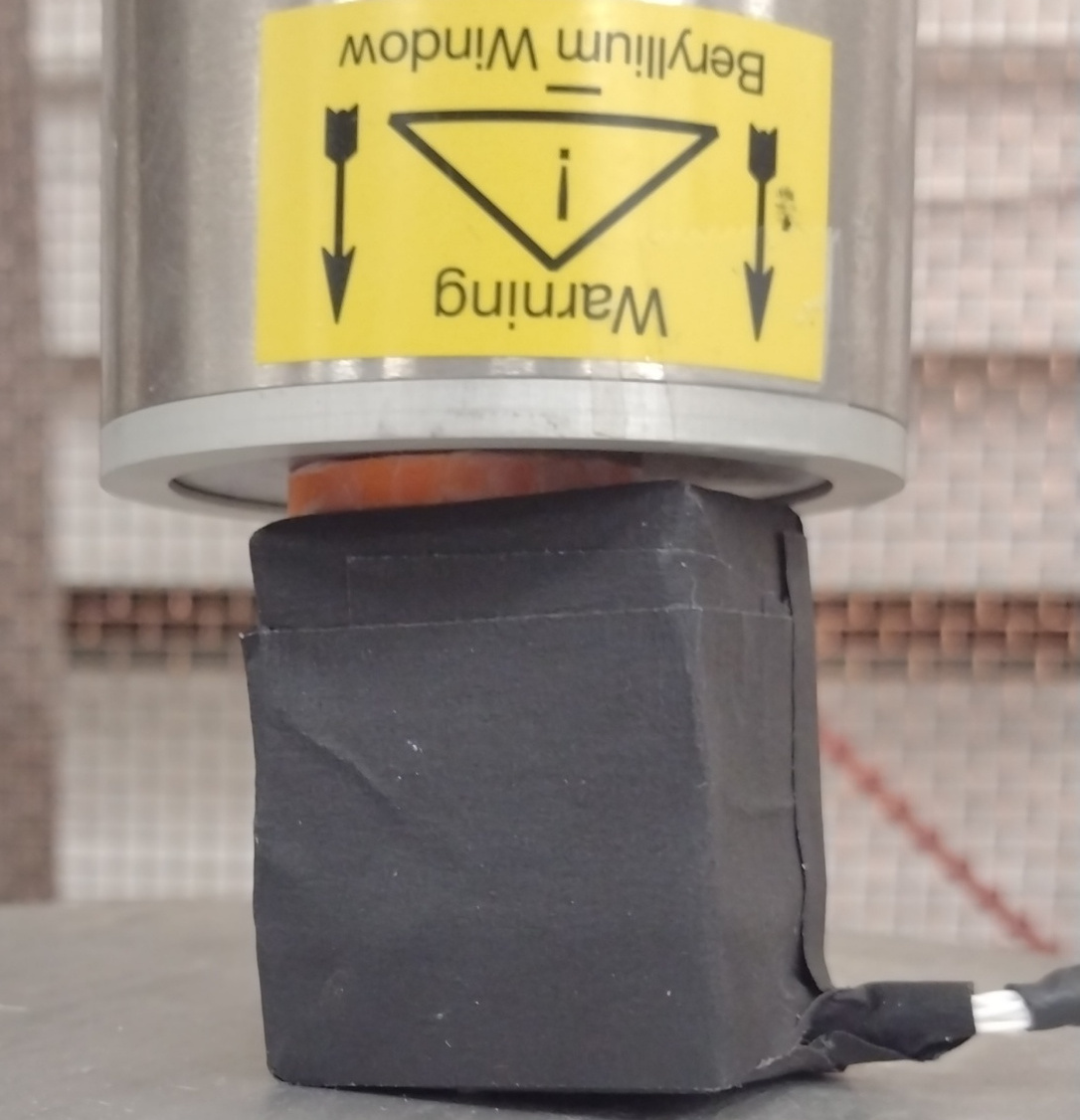}
\caption[Flight Detector and X-ray detector]{The flight detector (bottom) and the X-ray detector (top) with a source in between them.}
\label{fig:GAGGWithXray} 
\end{figure}

\begin{figure}[htbp]
\centering
\includegraphics[width = 0.6\textwidth]{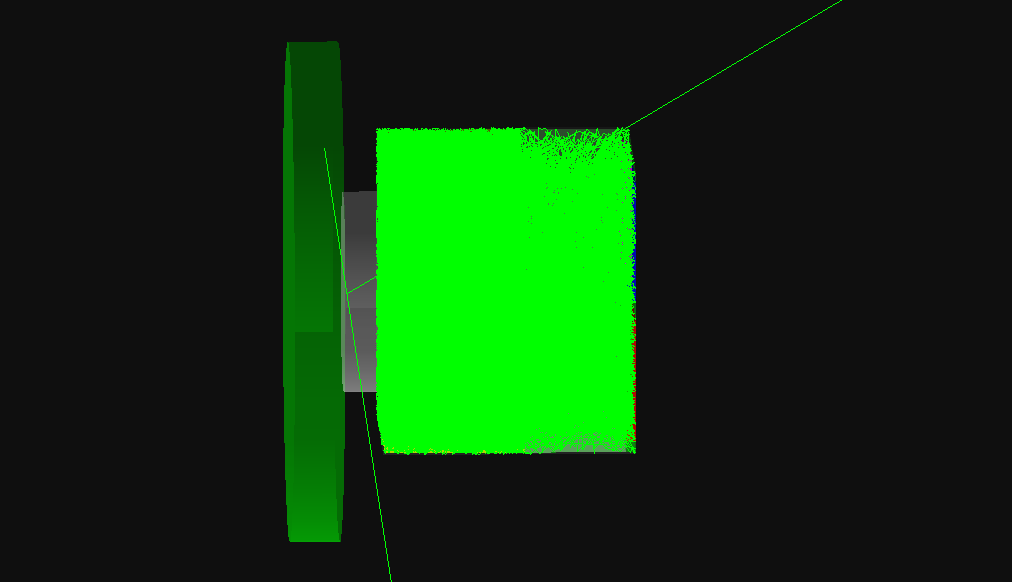}
\caption[Geant4 Na22 Decay]{A visualization of the Geant4 simulation used to calculate the efficiency of the GAGG detector. The green cylinder is the X-ray detector, the first 14mm of the large rectangular prism are the 4 GAGG crystals, then 10mm of quartz window with the 4 SiPMs on the end of the quartz, and finally the grey disk is the plastic source with a small deposit embedded at the end. Shown is a $^{22}$Na decay with a positron annihilation and gamma de-excitation.}
\label{fig:GEANT4 sim with Na22} 
\end{figure}

Lastly, a differential discriminator is used to trigger on the X-ray signal and save the waveforms for analysis (model is Ortec 583B CF Differential Discriminator). This allows us to select the relevant energy regime for triggering on the X-ray and prevent triggering on the vast majority of signals that would otherwise trigger an event. Note that the signal from the X-ray detector is split using an amplifier with one signal going to the differential discriminator. The trigger is set by determining the voltage values corresponding to the X-ray band, specifically by setting the differential discriminator to just above the absolute minimum setting and then lowering the upper threshold until the solid band was starting to disappear and then adjusting the upper level to just above that point, a standard technique for triggering on the photopeak of a low energy X-ray using a differential discriminator. The noise reduction was tested and applied with these discriminator values. For comparison of the simulated detector to the real lab setup, see Figures \ref{fig:GAGGWithXray} \& \ref{fig:GEANT4 sim with Na22}. For a view of the signals as seen by the oscilloscope (discussed in more detail later), see Figure \ref{fig:OscopeSignalZn65}.

We used an oscilloscope as our data acquisition method for both the GAGG prototype detector and the X-ray detector when triggered by the differential discriminator. For both detectors we took the integral of the waveform over time and the resulting value in Vns is binned into a histogram. The timing window of the oscilloscope is wide enough to include all of the trigger detector information and most of the waveform of the flight detector. This timing window is wide enough to include all of the events of the investigated topology and thus not significantly underestimate the  efficiency from missed expected events, but not too wide as to expect the integration method to cover many double pulses or to significantly overestimate the efficiency from coincident background events. It is worth noting that adding additional cuts to the data based on a time distribution (such as the time of maximum signal height) could potentially improve the noise rejection techniques as well as restricting the triggering topology to include events that are slightly more likely to be a valid coincident gamma as opposed to coincident. However, the real electronics for the flight detector will be using a wider timing window and an analysis on the mission data with a restriction to include only the timing window used in this work could be performed to characterize flight detector performance.

We then used ROOT \cite{rene_brun_2020_3895860} to generate histograms of these integrated signals and by integrating over the fit function corresponding to the histograms we determine the number of X-rays detected by the X-ray detector and the number of gamma rays detected by the prototype detector. Since the final value we are interested in is the measured vs. simulated photopeak efficiency, we are interested in the integral of the full energy deposition region of the signal. After fitting the histograms and taking the means of the Gaussians, a calibration curve (see Figures \ref{fig:CalibrationCurveMeasured} \& \ref{fig:CalibrationCurveSimulated}) is constructed using the known gamma energy values. Linearity of the calibration curve is both expected and seen.

\section{Simulations}\label{sec:Simulations}

\subsection{Simulation Overview}\label{sec:SimUlationOverview}
To measure the efficiency of the flight detector when compared to simulation, we used  simulations to determine the portion of gamma rays that were emitted by the source that actually interact with the detector to produce a signal. Simulations of the flight detector were done in Geant4 \cite{1610988} \cite{AGOSTINELLI2003250} \cite{ALLISON2016186} to determine the expected portion of emitted gamma rays.

The Geant4 simulation takes into account the scintillation yield as a function of wavelength for the scintillating material and optical photon counting in the SiPM including the quantum efficiency of the SiPM as a function of wavelength, though this is just implemented for future development. Note that both of these last properties are for future development and are perhaps not important for this paper. The relevant source is modeled as a small disk that emits gamma rays isotropically at each point of the disk and further embedded in a plastic disk with the standard dimensions of the Spectrum Technologies sources (1 inch in diameter and 0.125 inches in thickness). Modeling the disk source is negligible for most of the runs but was necessary for simulating the $^{22}$Na since the positrons need matter to annihilate with. See Figure \ref{fig:GEANT4 sim with Na22} for a view of the detector geometry.

Under these considerations, an interacting gamma ray in the simulated detector produces an optical photon count distribution that closely resembles the measured distribution of SiPM signal peaks in the real detector.

\subsection{Simulation of Flight Detector}
The simulated triggers were created using a logic surface that mimics the X-ray detector, this can be used to select for events that meet the criteria we choose. These simulated triggers were used to determine the correction factor (see Equation \ref{CorrectionFactorEquation}) for the $^{22}$Na signals and the $^{60}$Co signals. The number of expected photopeaks for $^{65}$Zn and $^{54}$Mn were calculated by simply spawning in an isotropic gamma ray source of the known energy, and fitting the resulting optical photon count histogram.
\begin{equation}
    \textbf{Correction Factor} = \frac{\textbf{Simulated Photopeaks Detected}}{\textbf{Simulated Gammas Emitted}}
    \label{CorrectionFactorEquation}
\end{equation}

\begin{equation}
    \textbf{Trigger Ratio} = \frac{\textbf{Number of Possible Triggers}}{\textbf{Number of Decays}}
    \label{TriggerRatioEquation}
\end{equation}

\begin{figure}[htbp]
\centering
\includegraphics[width = 0.9\textwidth]{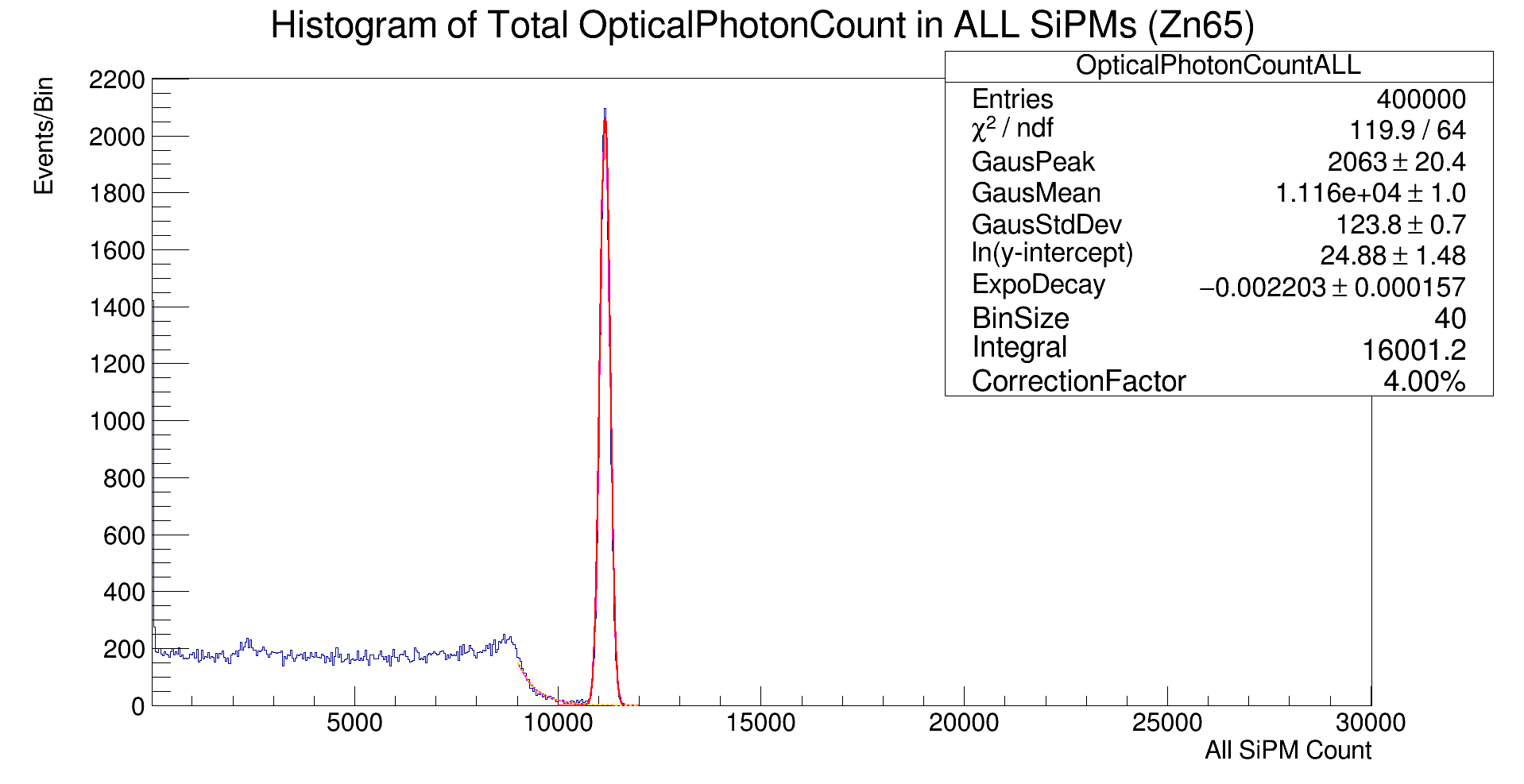}
\caption[Zn65 Simulated  Signals]{Simulated $^{65}$Zn 1115 keV gamma interacting with GAGG detector.}{\label{fig:Zn65 Sim}}
\end{figure}

The most important result from the simulation is the number of optical photons counted by the SiPMs. The optical photon count distribution looks like the standard gamma signal distribution from standard gamma ray spectroscopy experiments, with a broad Compton scattering region and a narrow full energy deposition region, see Figure \ref{fig:Zn65 Sim}. We use ROOT to fit a Gaussian function to the histogram of optical photons, and use the fit parameters to determine the number of gammas that are expected to be seen with a full energy deposit in the real detector. The portion of full energy deposits out of the total gammas produced is dubbed the correction factor. Note that this is without the trigger ratio as that is kept as a separate factor for consistency.

\begin{table}[htbp]
    \centering
        \caption{The correction factors obtained by Geant4 simulation along with the relevant trigger ratios for the interaction considering the trigger. (Values over 99\% are considered 100\% and the value of 200\% for $^{22}$Na is due to there being 2 annihilation gammas to potentially trigger on for each emitted 1274 keV gamma).}
    \begin{tabular}{llll}
        \textbf{Source} & \textbf{Energy}&\textbf{CF} & \textbf{BR} \\\hline\hline
        $^{65}$Zn & 1115 keV & 4.00\% & 50.23\% \\\hline
        $^{54}$Mn & 835 keV & 5.60\% & 100\% \\\hline
        $^{22}$Na & 1274 keV & 2.03\% & 200\% \\\hline 
        $^{22}$Na & 1785 keV & 0.69\% & 100\% \\\hline 
        $^{60}$Co & 1173 keV & 3.79\% & 100\% \\\hline 
        $^{60}$Co & 1332 keV & 3.28\% & 100\% \\\hline       
    \end{tabular}
    \label{tab:CorrectionFactors}
\end{table}

\begin{figure}[htbp]
\centering
\includegraphics[width = 0.9\textwidth]{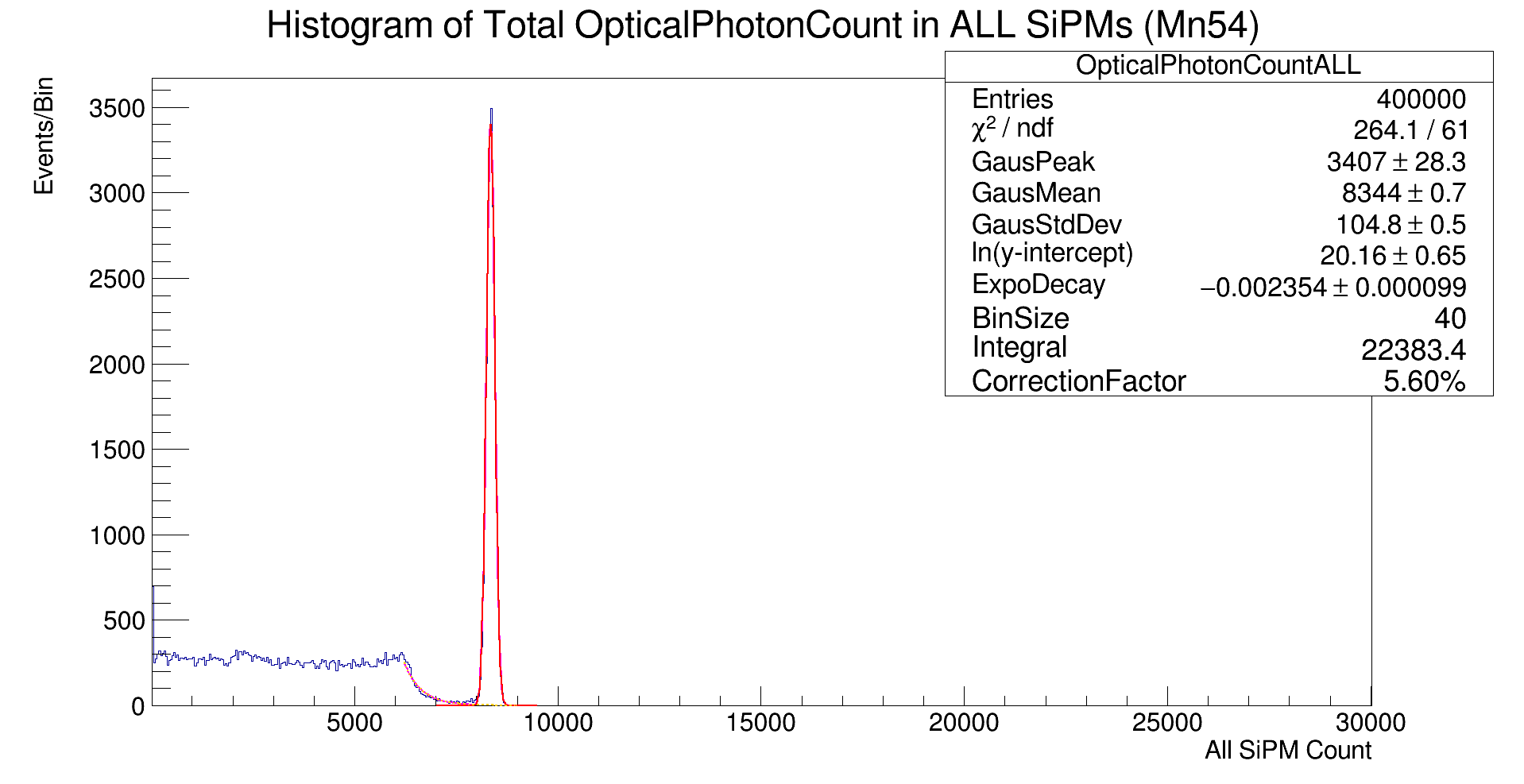}
\caption[Mn54 Simulated Signals]{Simulated $^{54}$Mn 835 keV gamma interacting with GAGG detector.}{\label{fig:Mn54 Sim}}
\end{figure}

\begin{figure}[htbp]
\centering
\includegraphics[width = 0.9\textwidth]{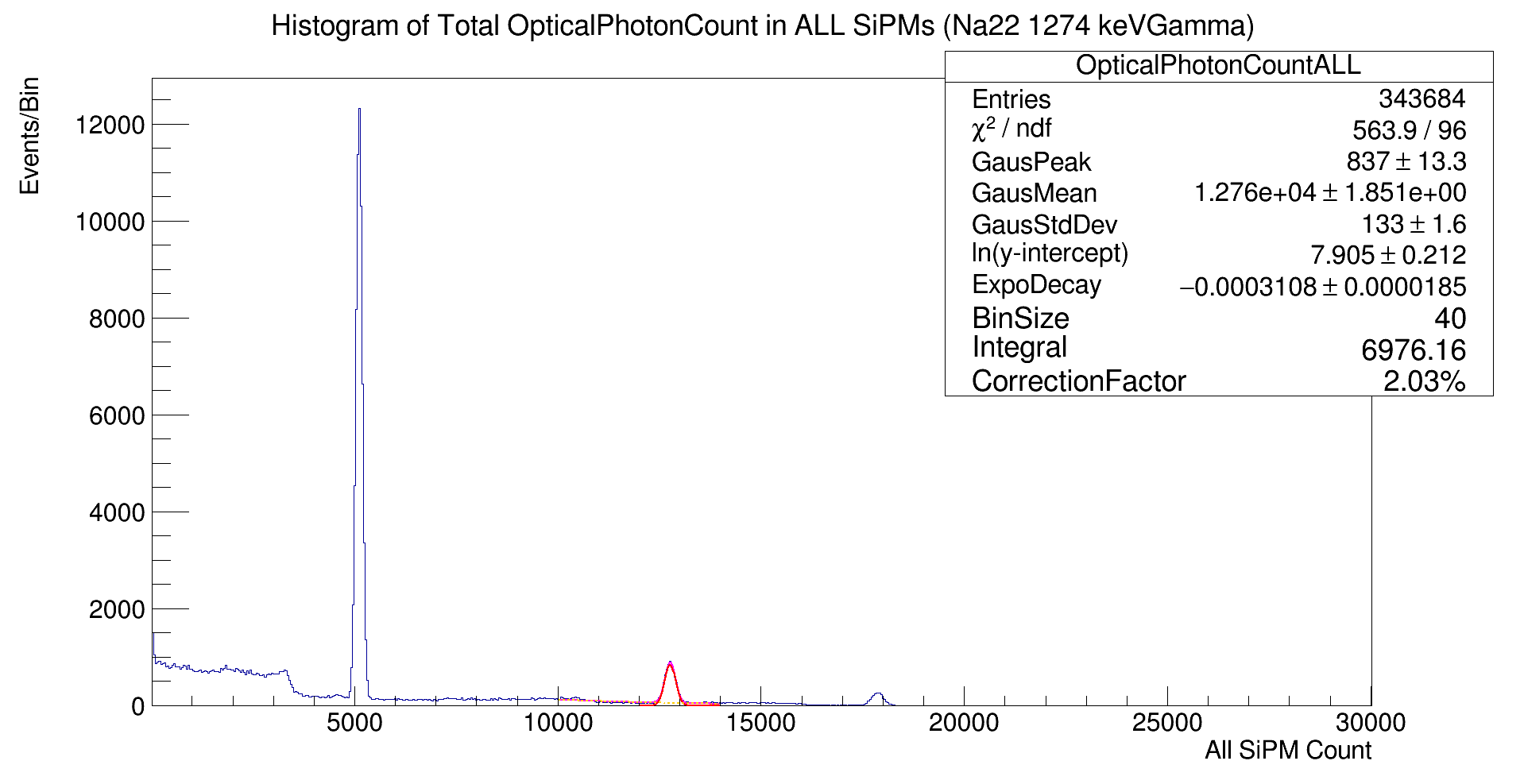}
\caption[Na22 Simulated Gamma Signals]{Simulated $^{22}$Na decays with X-ray logic triggering on one of the 511 keV annihilation gammas. The fit is for the 1274 keV gamma interacting with the GAGG detector.}{\label{fig:Na22 Gamma Sim}}
\end{figure}

\begin{figure}[htbp]
\centering
\includegraphics[width = 0.9\textwidth]{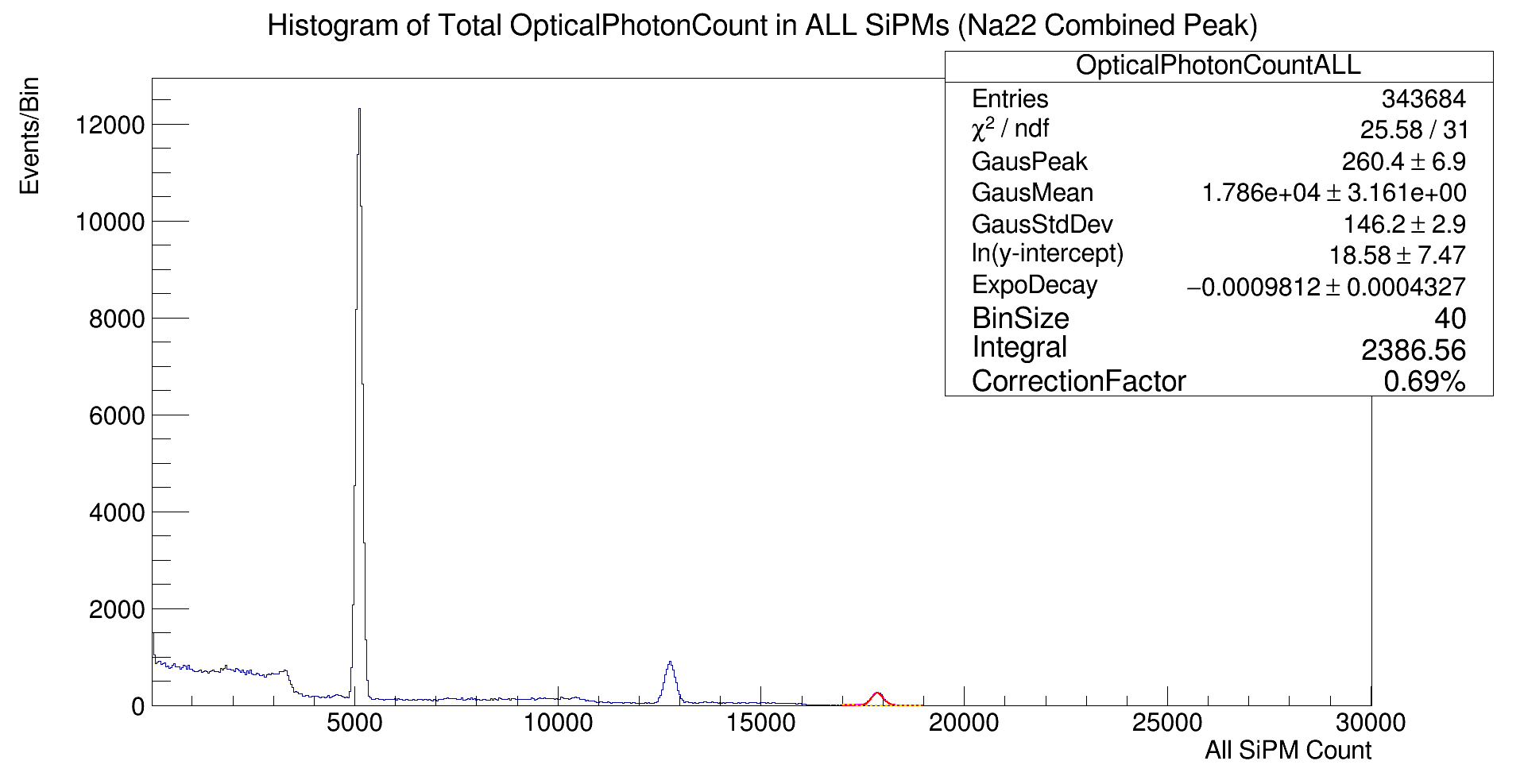}
\caption[Na22 Simulated Combined Signals]{Simulated $^{22}$Na decays with X-ray logic triggering on one of the 511 keV annihilation gammas. The fit is for the combined peak of both the 1274 keV gamma and one of the 511 keV annihilation gammas.}{\label{fig:Na22 Combined Sim}}
\end{figure}

\begin{figure}[htbp]
\centering
\includegraphics[width = 0.9\textwidth]{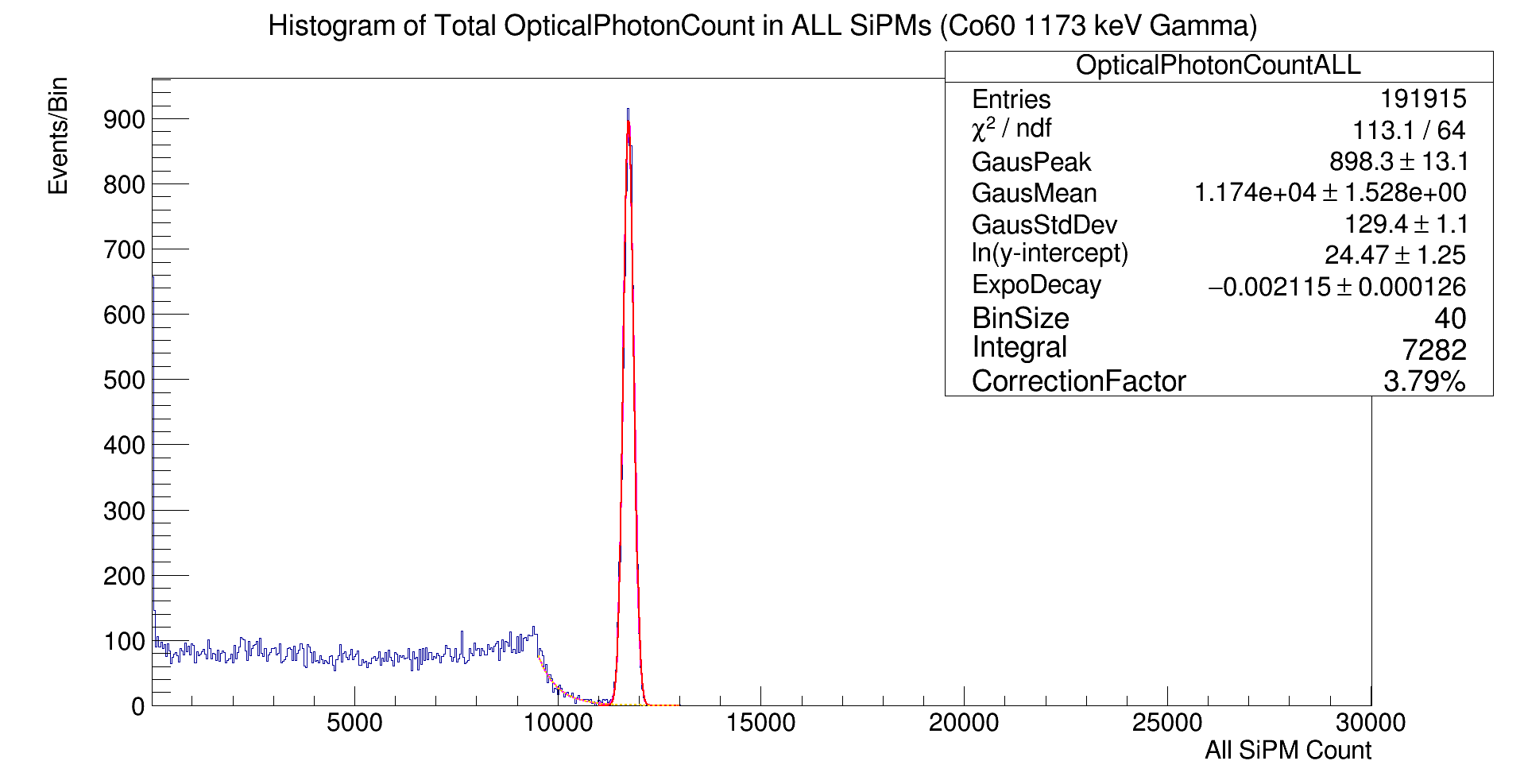}
\caption[Co60 Simulated Signals]{Simulated $^{60}$Co 1173 keV decays with X-ray logic triggering on the 1332 keV gamma.}{\label{fig:Co60 Gamma1 sim}}
\end{figure}

\begin{figure}[htbp]
\centering
\includegraphics[width = 0.9\textwidth]{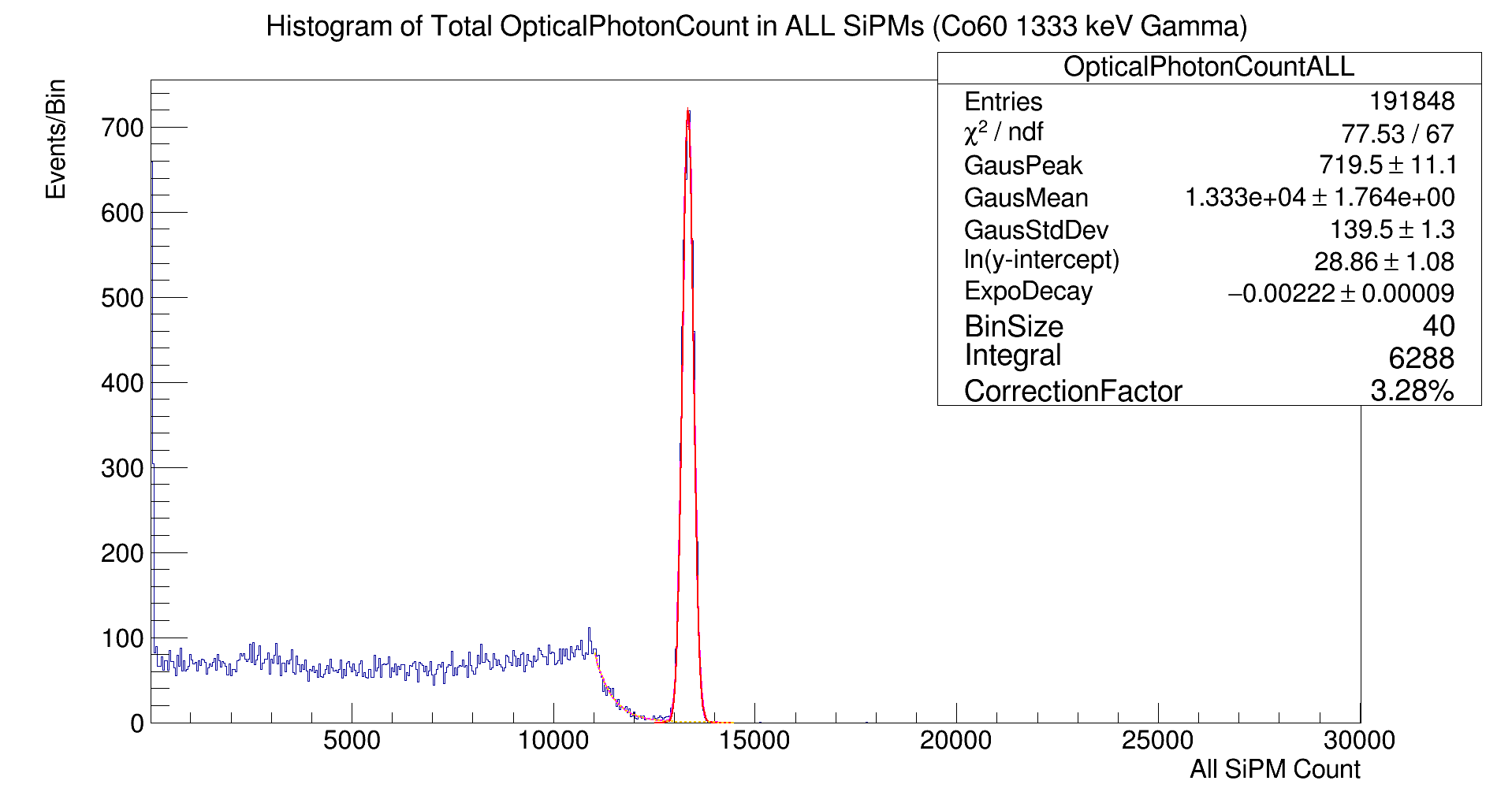}
\caption[Co60 Simulated Signals 2]{Simulated $^{60}$Co 1332 keV decays with X-ray logic triggering on the 1173 keV gamma.}{\label{fig:Co60 Gamma2 sim}}
\end{figure}

A summary of the results of these simulations can be found in Table \ref{tab:CorrectionFactors}.

\section{Results}\label{Results}

\subsection{Definition of Efficiency}

We are using the term measured vs. simulated photopeak efficiency to mean the ratio of gamma rays that produce a full energy deposit in the detector to gamma rays that were expected to produce a full energy deposit in the detector according to simulations.

\begin{equation}
    \textbf{Efficiency} = \frac{\textbf{Detected}}{\textbf{Expected}}
\end{equation}

In this context, the number of detected gamma rays is the integral over the full energy capture photopeak that is characteristic of the known energy gammas that we are looking for. The number of expected gamma rays is given in Equation \ref{ExpectedEquation}.

\begin{equation}
    \textbf{Expected} = \textbf{CorrectionFactor} \cdot \textbf{TriggerRatio} \cdot \textbf{Triggers}
    \label{ExpectedEquation}
\end{equation}

\subsection{Results for Flight Detector using X-ray-Triggered Gamma Sources}
\begin{figure}[htbp]
\centering
\includegraphics[width = 0.9\textwidth]{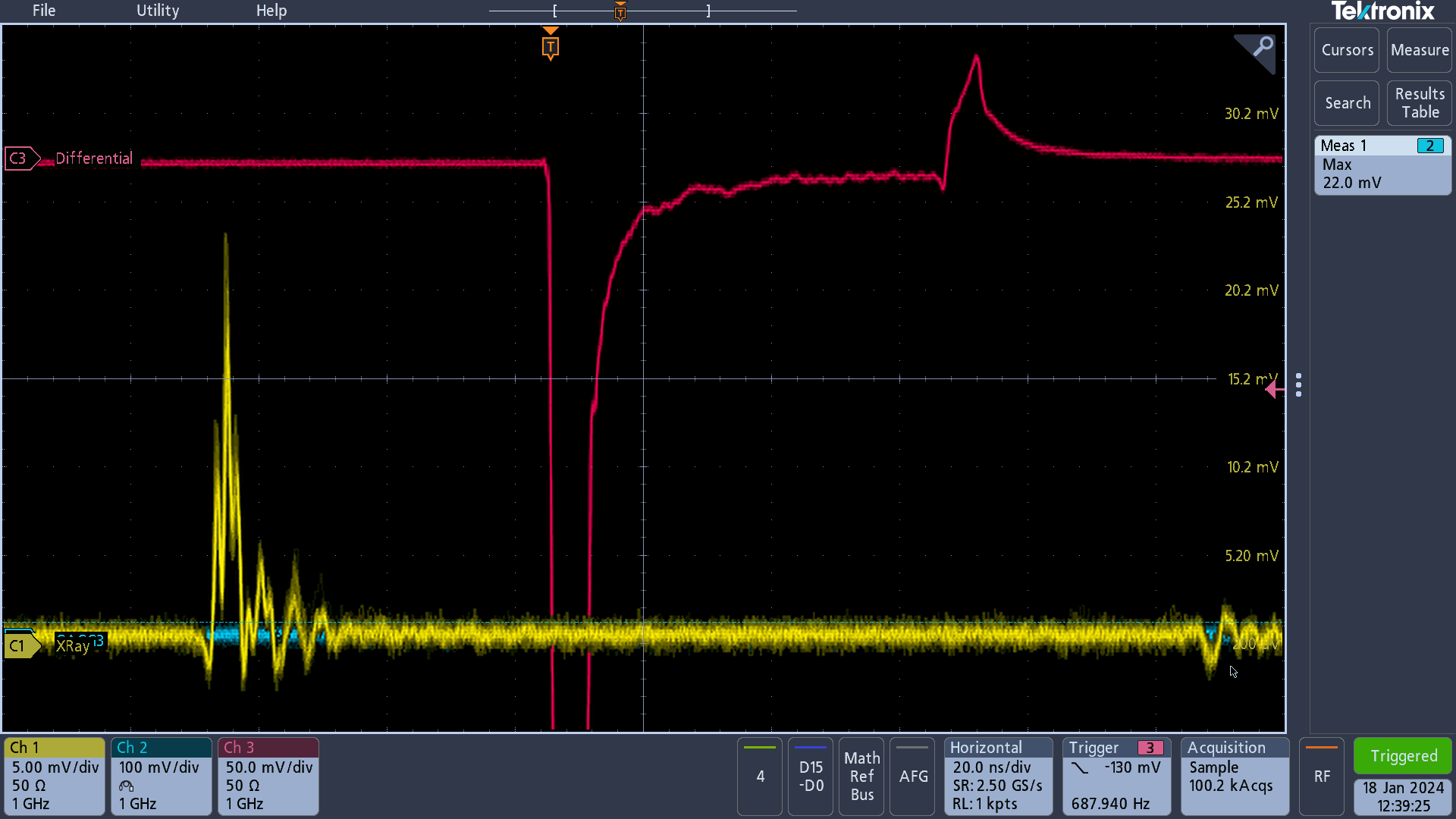}
\caption[Noise Events]{Yellow: X-ray detector signals. Red: Differential discriminator signal used as the trigger. Blue: Flight detector signals. View of oscilloscope screen when triggering on the differential discriminator with no source. The noise events have a mostly consistent form and are much shorter in time than signals.}{\label{fig:OscopeSignalNoise}}
\end{figure}

\begin{figure}[htbp]
\centering
\includegraphics[width = 0.9\textwidth]{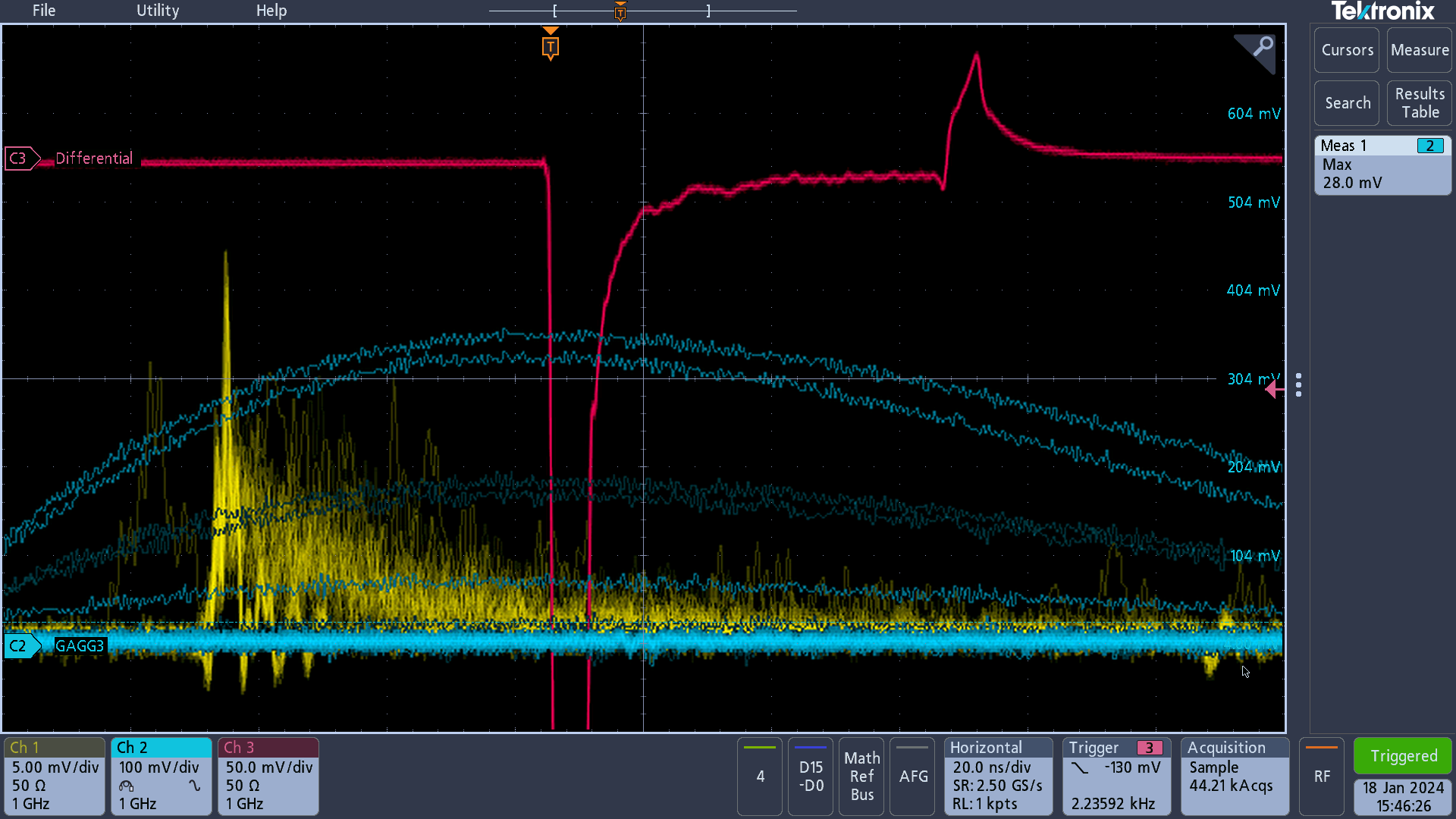}
\caption[Zn65 Events]{Yellow: X-ray detector signals. Red: Differential discriminator signal used as the trigger. Blue: Flight detector signals. View of oscilloscope screen when triggering on the differential discriminator with the $^{65}$Zn source in position. The 2 blue traces that peak at just above 304 mV are presumably from the 1.115 MeV gammas.}{\label{fig:OscopeSignalZn65}}
\end{figure}

\begin{figure}[htbp]
\centering
\includegraphics[width = 0.9\textwidth]{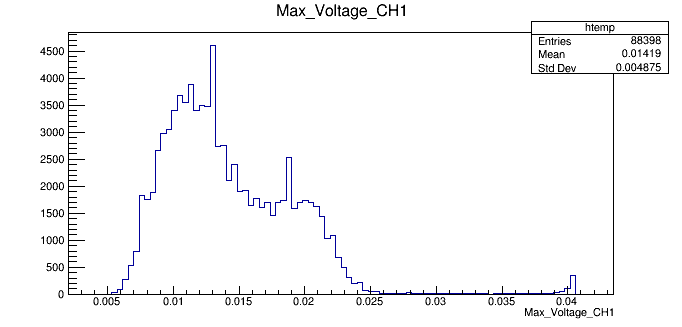}
\caption[Zn65 data without trimming]{Data from $^{65}$Zn before trimming.}
\label{fig:Untrimmed} 
\end{figure}

\begin{figure}[htbp]
\centering
\includegraphics[width = 0.9\textwidth]{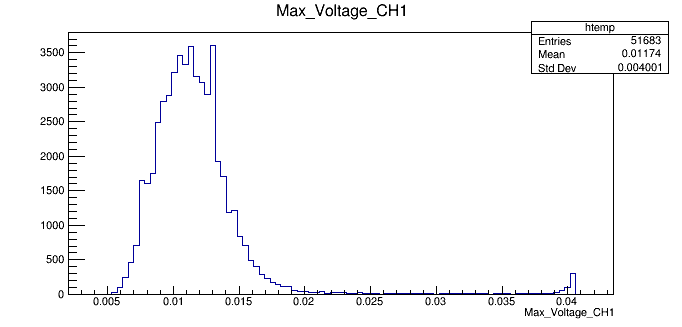}
\caption[Zn65 data with trimming]{Data from $^{65}$Zn after trimming. Note that the noise and the real signals are distinct.}
\label{fig:Trimmed} 
\end{figure}

\begin{figure}[htbp]
\centering
\includegraphics[width = 0.9\textwidth]{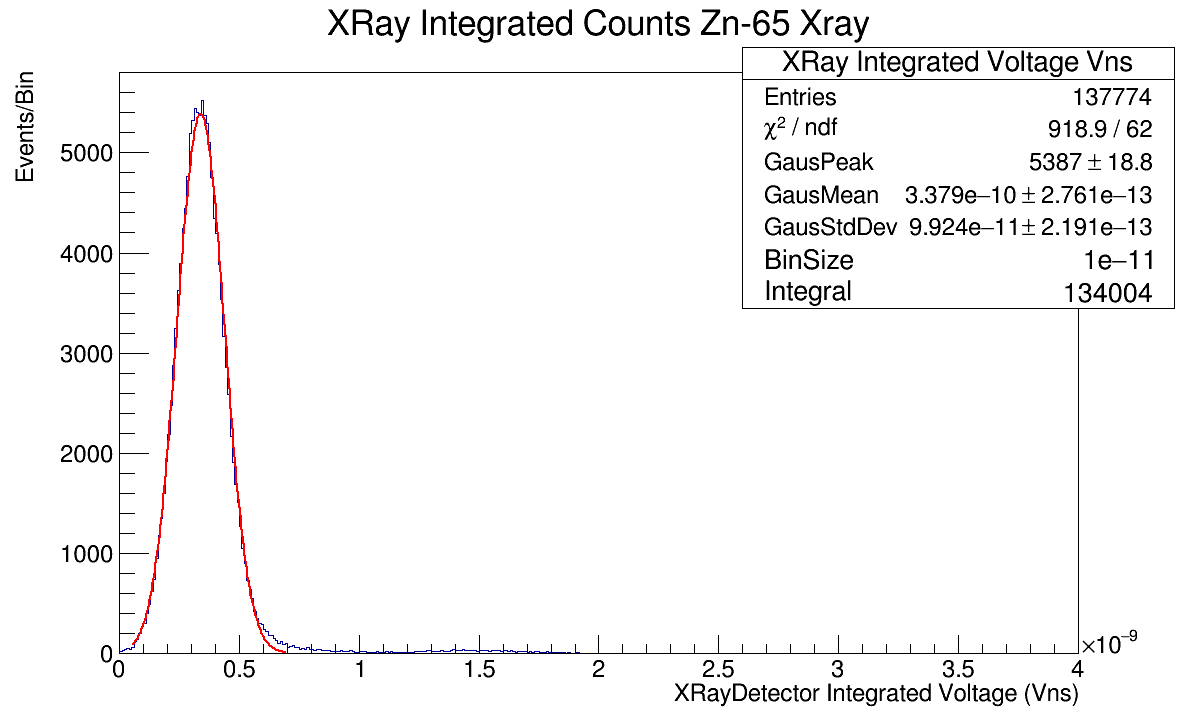}
\caption[Zn65 Xray Signals]{The 8 keV X-ray signals from $^{65}$Zn in the X-ray detector.}{\label{fig:Zn65Xray}}
\end{figure}

\begin{figure}[htbp]
\centering
\includegraphics[width = 0.9\textwidth]{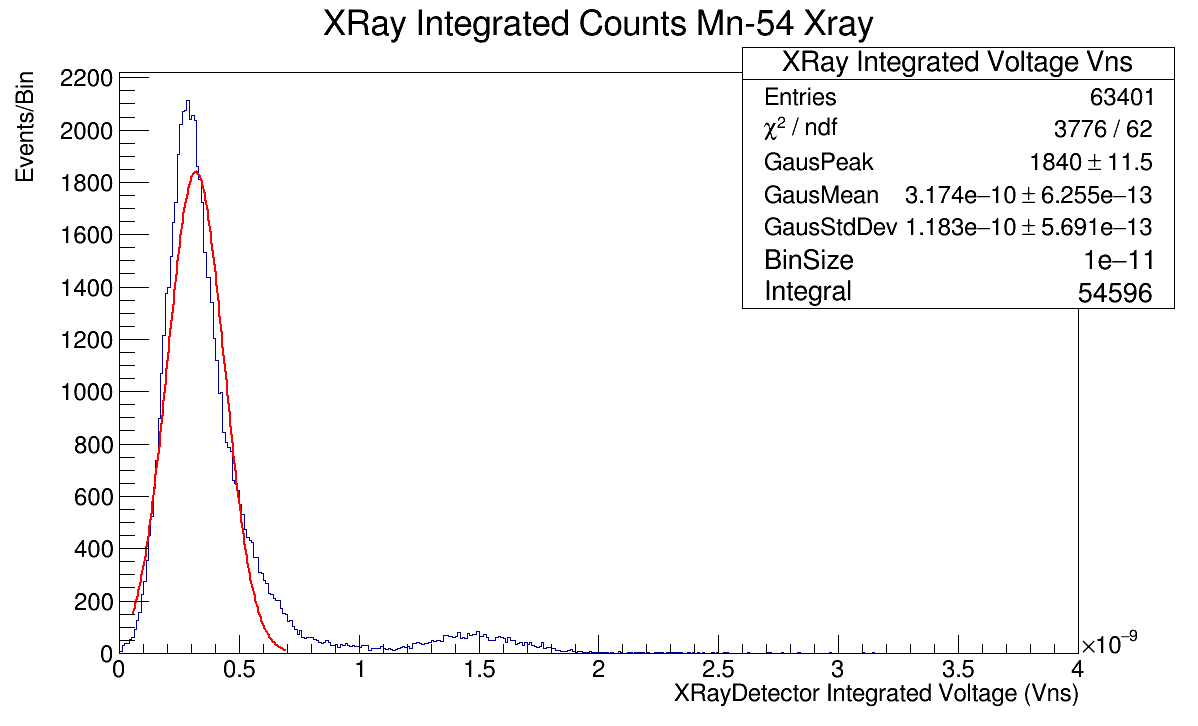}
\caption[Mn54 Xray Signals]{The 5.5 keV X-ray signals from $^{54}$Mn in the X-ray detector. Note that the detector is rated for as low as 5 keV, so more noise and backgrounds are to be expected}{\label{fig:Mn54Xray}}
\end{figure}

For the flight detector with real sources we get results in the form of histograms of the integrated signals from both the X-ray detector and the flight detector. There are several important considerations to take into account to interpret these data. For the electron capture sources, $^{65}$Zn and $^{54}$Mn, the X-rays are low energy and therefore the signals are of similar size to a noise signal that is generated by the X-ray detector and/or the electronics. However, we can use a further discrimination technique in software to remove these noise events. Specifically, anytime a signal rises above a relatively low value but then crosses the voltage axis at zero within 20 ns, the event is removed from the data set. The noise removal technique applied to a run with no sources but triggering using the differential discriminator brought the noisy data set down from 3827 events down to just 32, a 99.2\% reduction in noise events, see Figures \ref{fig:OscopeSignalNoise} - \ref{fig:Trimmed}. The resultant histograms formed from noise reduction are below, see Figures \ref{fig:Zn65Xray} \& \ref{fig:Mn54Xray}. After noise reduction, the leftover data is fit with a Gaussian (the previous data was for a trial run specifically to determine the efficacy of the noise removal procedure). The Gaussian is integrated to determine the number of X-rays we are using as triggers.
Note that the possibility of triggering on surviving noise events could artificially lower the measured efficiency of the flight detector by inflating the number of expected gamma rays. To address this, the rates of noise to real triggers combined with the noise reduction technique can be used to estimate the number of extra triggers from the X-ray detector. The rate of triggering with no source is roughly 688 Hz whereas the rate with the $^{65}$Zn source is roughly 2236 Hz, or a rate of 1548 Hz for valid triggers (see Figures \ref{fig:OscopeSignalNoise} \& \ref{fig:OscopeSignalZn65}). 0.8\% of the 688 Hz noise is roughly 5.5 Hz, or only 0.36\% of the valid triggering rate. Also the noise events form a very distinct and consistent shape whose integral hardly overlaps with the integral of the X-ray signals from $^{65}$Zn and $^{54}$Mn, further reducing the portion of the X-rays that are over-determined by noise. Due to the small size of this effect, and the fact that it was unique to only  $^{65}$Zn and $^{54}$Mn, it was ignored for the calculation of uncertainties.

\begin{figure}[htbp]
\centering
\includegraphics[width = 0.9\textwidth]{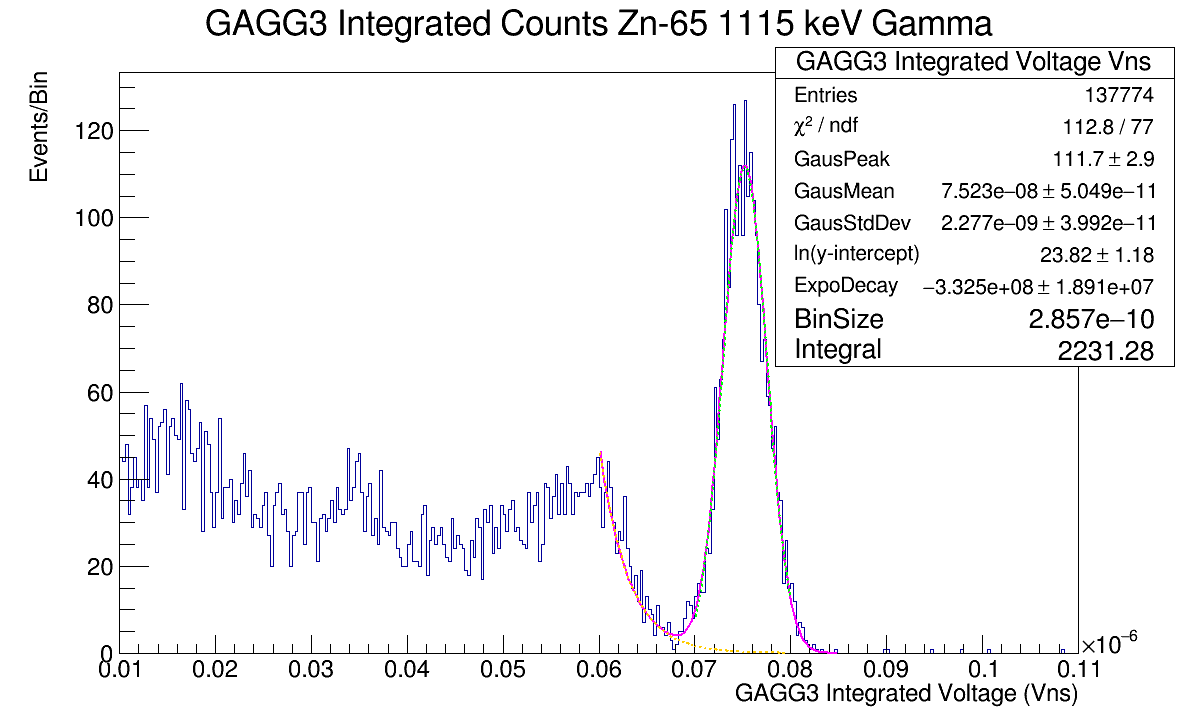}
\caption[Zn65 GAGG Signals]{The 1115 keV gamma ray signals from $^{65}$Zn in the flight detector.}{\label{fig:Zn65GAGG}}
\end{figure}

For the case of $^{65}$Zn, we saw 134004 X-rays in the X-ray detector. Using the trigger ratio and the correction factor obtained from Geant4, we expected to see 2692 gamma rays in the GAGG detector. The integral over the GAGG detector data gave 2231 gamma rays, thus a measured vs. simulated efficiency of 82.9\%.

\begin{figure}[htbp]
\centering
\includegraphics[width = 0.9\textwidth]{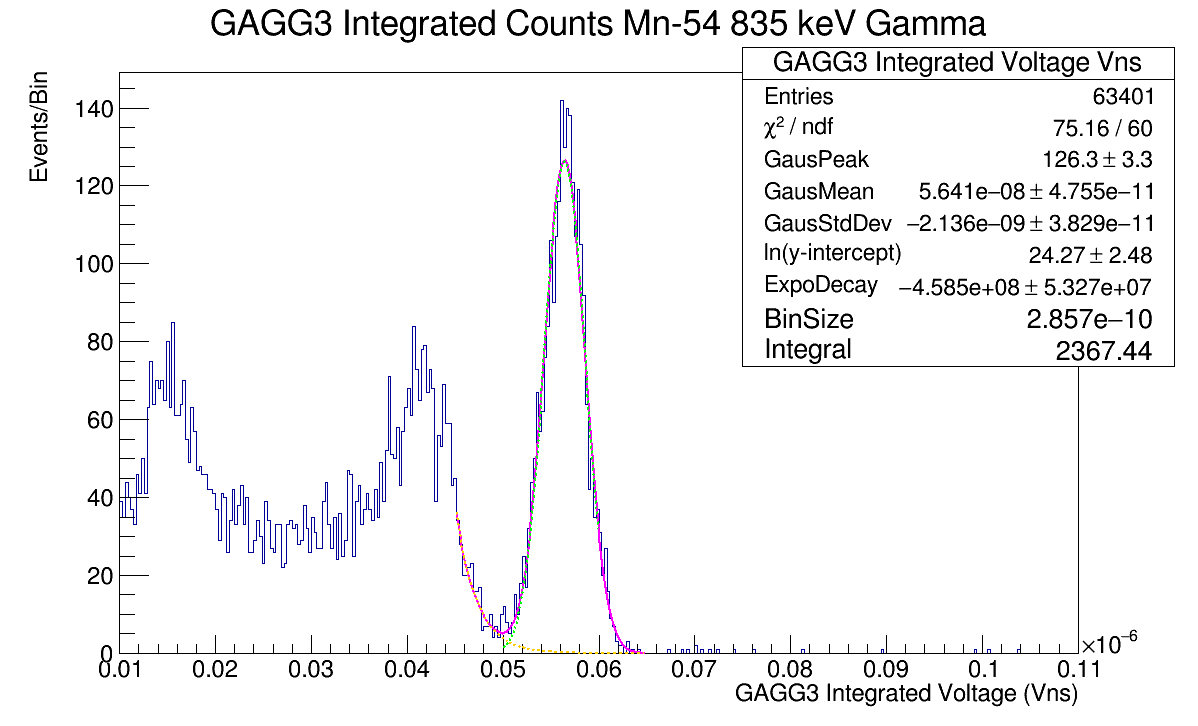}
\caption[Mn54 GAGG Signals]{The 835 keV gamma ray signals from $^{54}$Mn  in the flight detector.}{\label{fig:Mn54GAGG}}
\end{figure}

For $^{54}$Mn, we saw 54596 X-rays in the X-ray detector using the integral over the Gaussian. Using the correction factor of 5.60\%, we expected to see 3057 gammas in the GAGG detector. The integrated GAGG signal gave 2367 gammas, thus the measured vs. simulated efficiency is 77.4\%

\subsection{Results for Flight Detector using Annihilation-Triggered Gamma Sources}

\begin{figure}[htbp]
\centering
\includegraphics[width = 0.9\textwidth]{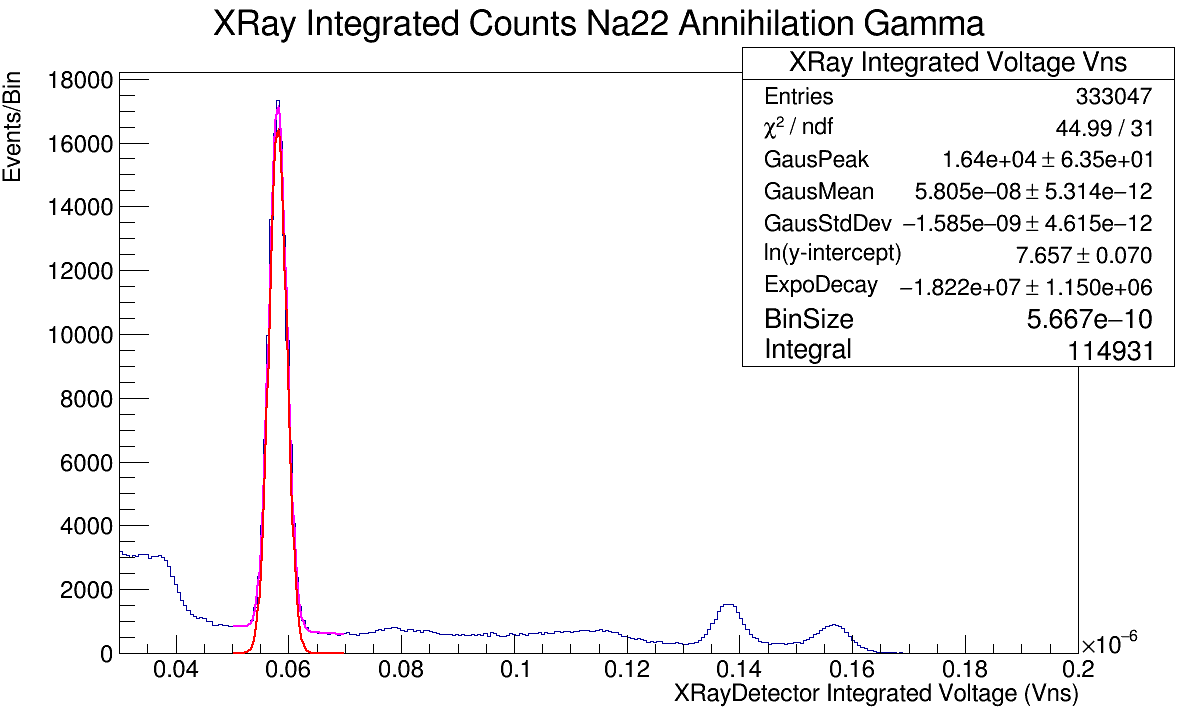}
\caption[Na22 X-ray Signals]{The 511 keV annihilation signal from $^{22}$Na in the X-ray detector.}{\label{fig:Na22AnnihilationXray}}
\end{figure}

\begin{figure}[htbp]
\centering
\includegraphics[width = 0.9\textwidth]{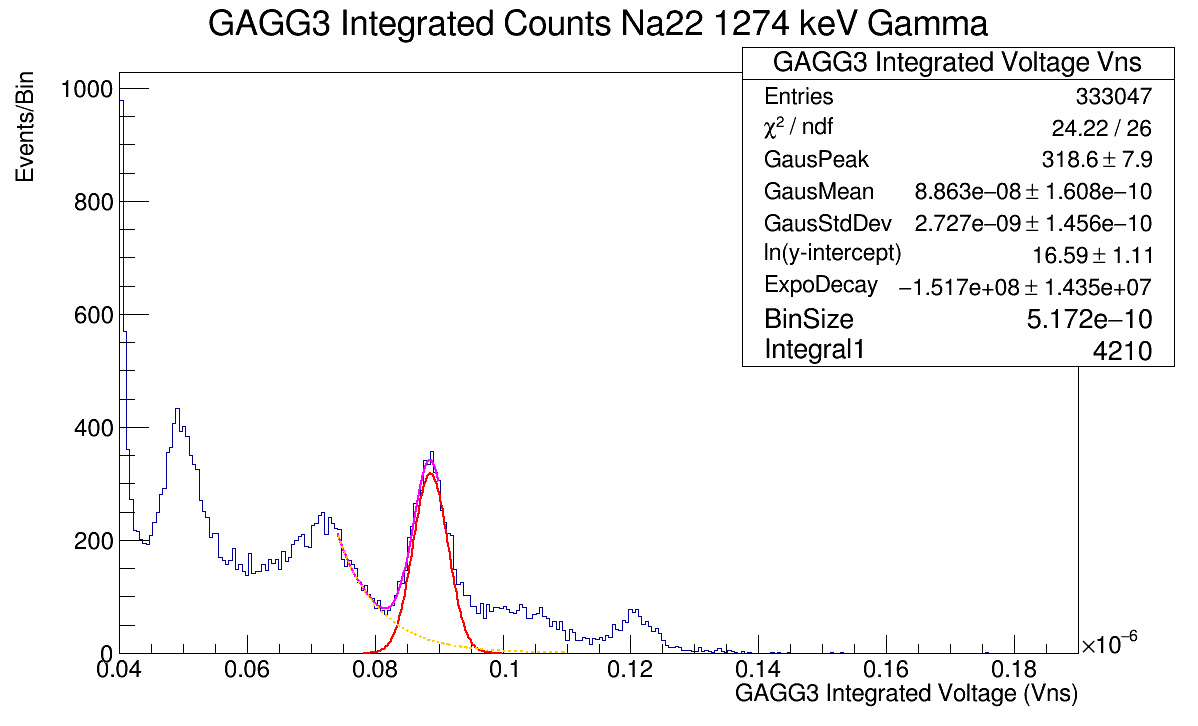}
\caption[Na22 GAGG Gamma Signals]{The 1274 keV de-excitation gamma signal from $^{22}$Na in the flight detector.}{\label{fig:Na22GammaGAGG}}
\end{figure}

\begin{figure}[htbp]
\centering
\includegraphics[width = 0.9\textwidth]{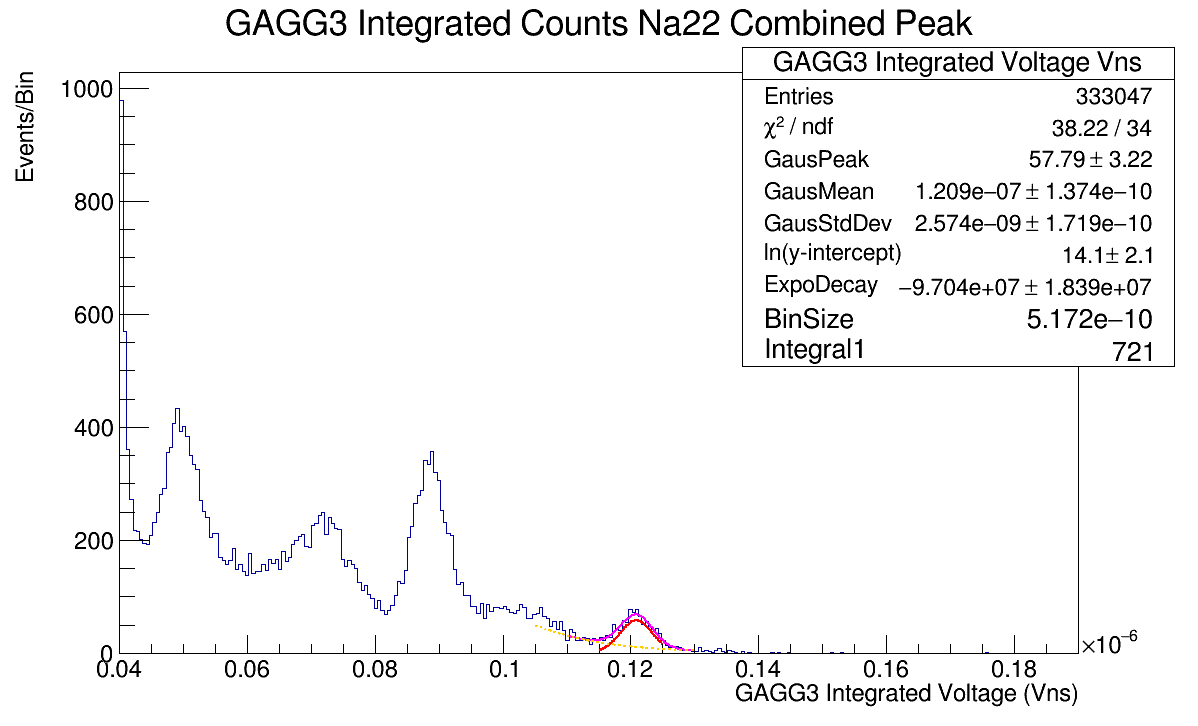}
\caption[Na22 GAGG Combined Signals]{The 1785 keV combined peak signal from both the annihilation and de-excitation gammas of $^{22}$Na in the flight detector.}{\label{fig:Na22CombinedGammaGAGG}}
\end{figure}

For $^{22}$Na, we saw 114931 positron annihilation gammas in the X-ray detector using the integral over the Gaussian. The trigger ratio for the 1274 keV nuclear de-excitation gamma is actually 200\% in this context since there is a 100\% chance to emit a gamma with respect to the positron emission and each annihilation produces 2 gammas. Thus, using the correction factor of 2.03\%, we expected to see 4666 gammas in the GAGG detector. The integrated GAGG signal gave 4210 gammas, thus the efficiency is 90.2\%.
In the case of $^{22}$Na, there is also the case where both the 1274 keV de-excitation gamma and one of the 511 keV positron annihilation gammas interact in the detector with a full energy deposit.
The correction factor for this case is 0.69\% and thus the predicted number of combined gamma signals is 793. We saw 721 gammas in the GAGG, or an efficiency of 90.9\%.

\begin{figure}[htbp]
\centering
\includegraphics[width = 0.9\textwidth]{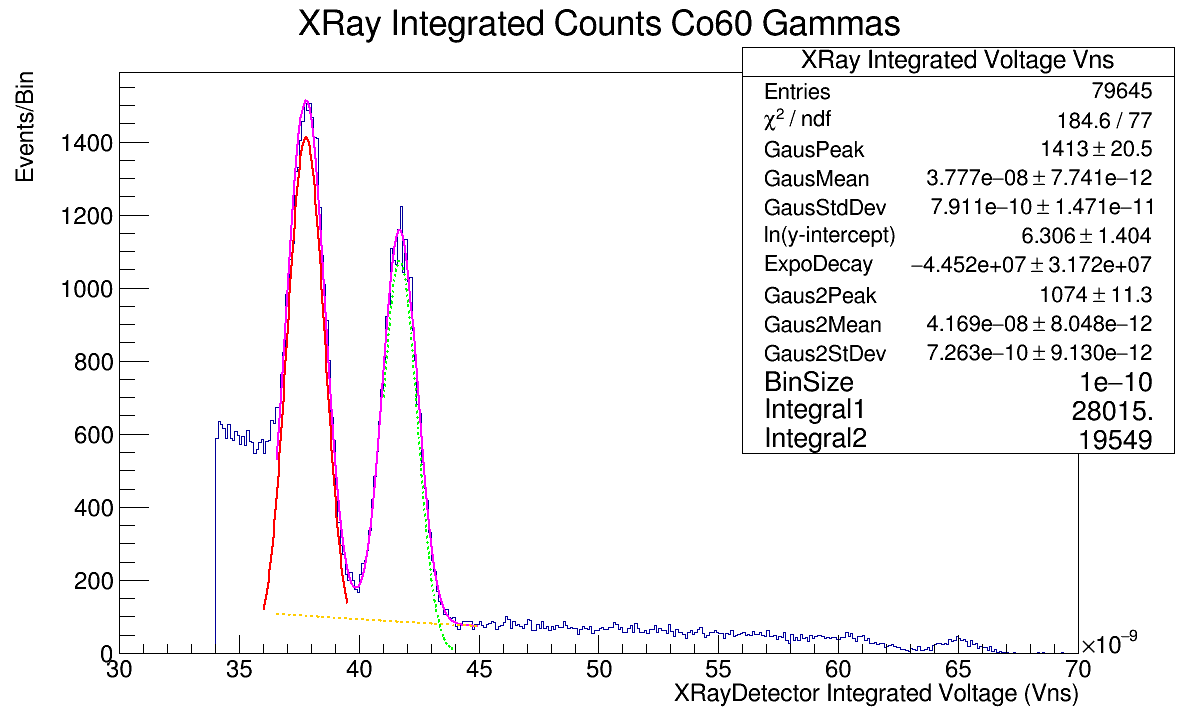}
\caption[Co60 X-ray Detector]{The $^{60}$Co gamma peaks from the X-ray Detector.}{\label{fig:Co60GammasX-rayDetector}}
\end{figure}

\begin{figure}[htbp]
\centering
\includegraphics[width = 0.9\textwidth]{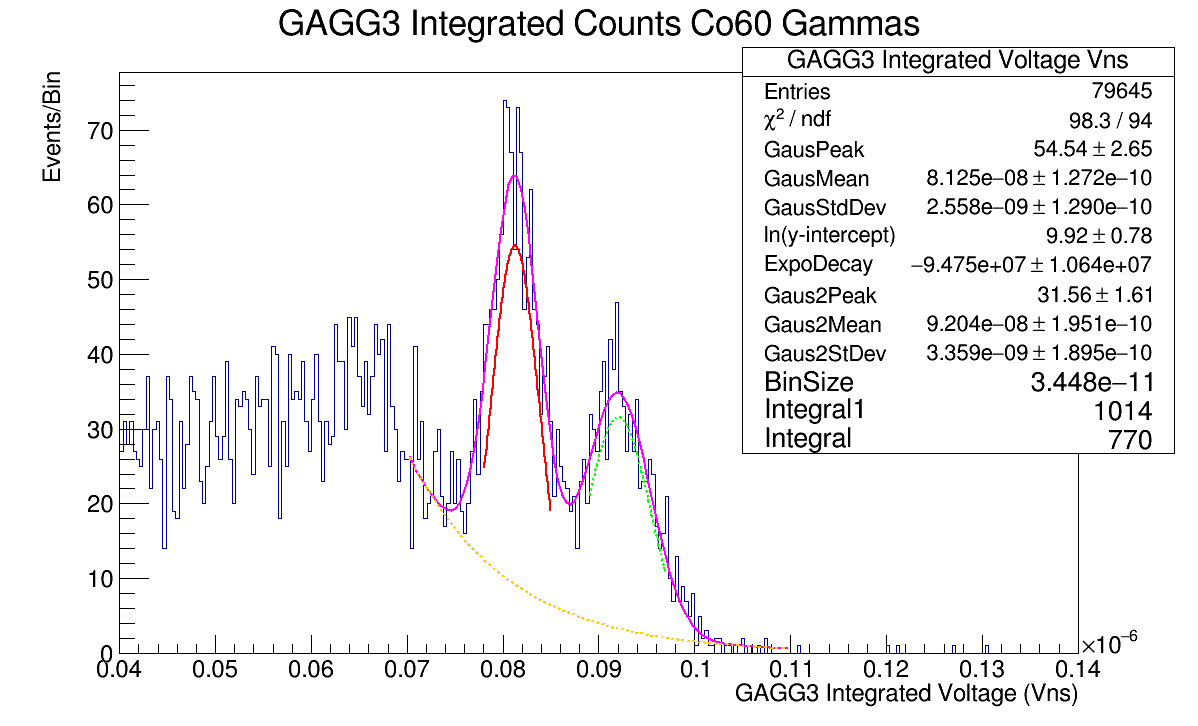}\caption[Co60 GAGG]{$^{60}$Co gamma peaks from the GAGG flight detector. This is the unrestricted data set.}{\label{fig:Co60GammasGAGGDetector}}
\end{figure}

\subsection{Results for Flight Detector Using a Gamma Cascade}

$^{60}$Co has 2 close gamma rays (hereafter referred to as gamma 1 and gamma 2, since both the energy and position in the cascade are ordered in an increasing fashion), so care must be taken not to trigger on the Compton edge of gamma 2 when looking for gamma 2 using gamma 1 as a trigger. The full data set is included as well as data sets that have been cut in software to select for only the appropriate region of the data.

We used the uncut data to determine the number of X-rays and therefore the number of expected gammas. From the double Gaussian fit, we saw 28015 of gamma 1 and 19549 of gamma 2 detected in the X-ray detector. From this we expected 740 of gamma 1 and 919 of gamma 2 in the flight detector.

Naively using the uncut data, we saw 1014 of gamma 1 and 770 of gamma 2. This is clearly suspicious since this would have made the detector over 100\% efficient for the detection of gamma 1. However, using the cut data sets, we saw only 676 gamma 1 and 761 gamma 2, meaning an efficiency 91.4\% and 82.8\% respectively.

\begin{figure}[htbp]
\centering
\includegraphics[width = 0.9\textwidth]{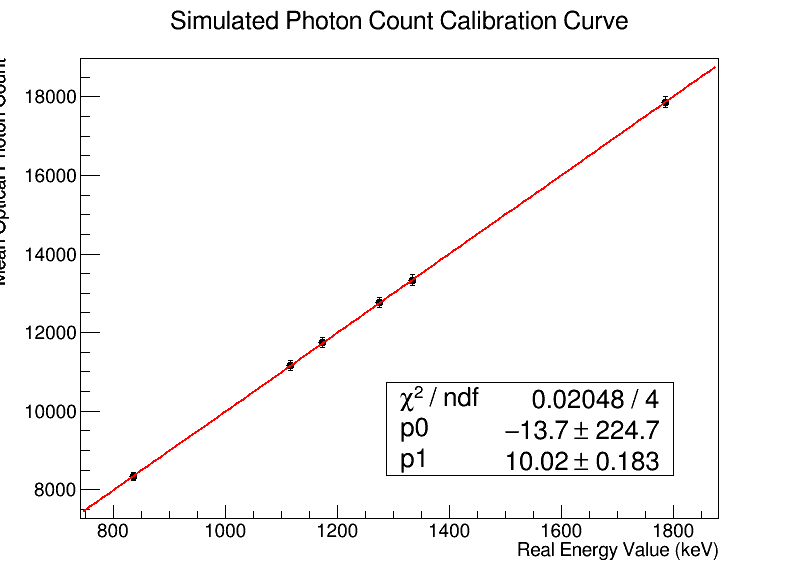}
\caption[Calibration Curve of Simulation Optical Photon Counts]{Calibration curve showing the simulated optical photon count vs. gamma energy. The first fit parameter is the intercept and the second is the slope.}
\label{fig:CalibrationCurveSimulated} 
\end{figure}

\begin{figure}[htbp]
\centering
\includegraphics[width = 0.9\textwidth]{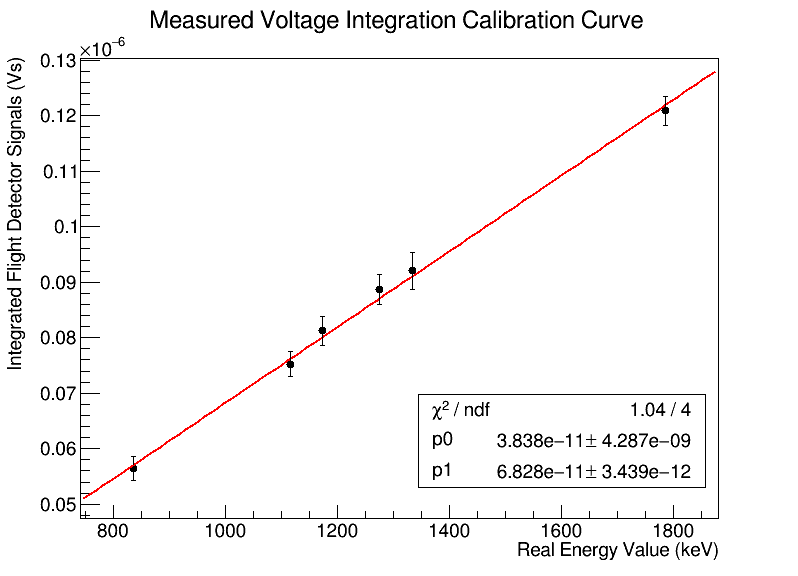}
\caption[Calibration Curve of Measured Integrated Signals]{Calibration curve showing the optical photon count vs. gamma energy for the GAGG flight detector. The first fit parameter is the intercept. The second is the slope.}
\label{fig:CalibrationCurveMeasured} 
\end{figure}

For a comparison of the simulated optical photon count calibration curve with the measured integrated voltage calibration curve see Figures \ref{fig:CalibrationCurveSimulated} \& \ref{fig:CalibrationCurveMeasured}. The main takeaway is that we are firmly in the linearity region of light detection for both the simulated and real detectors.

\begin{figure}[htbp]
\centering
\includegraphics[width = 0.9\textwidth]{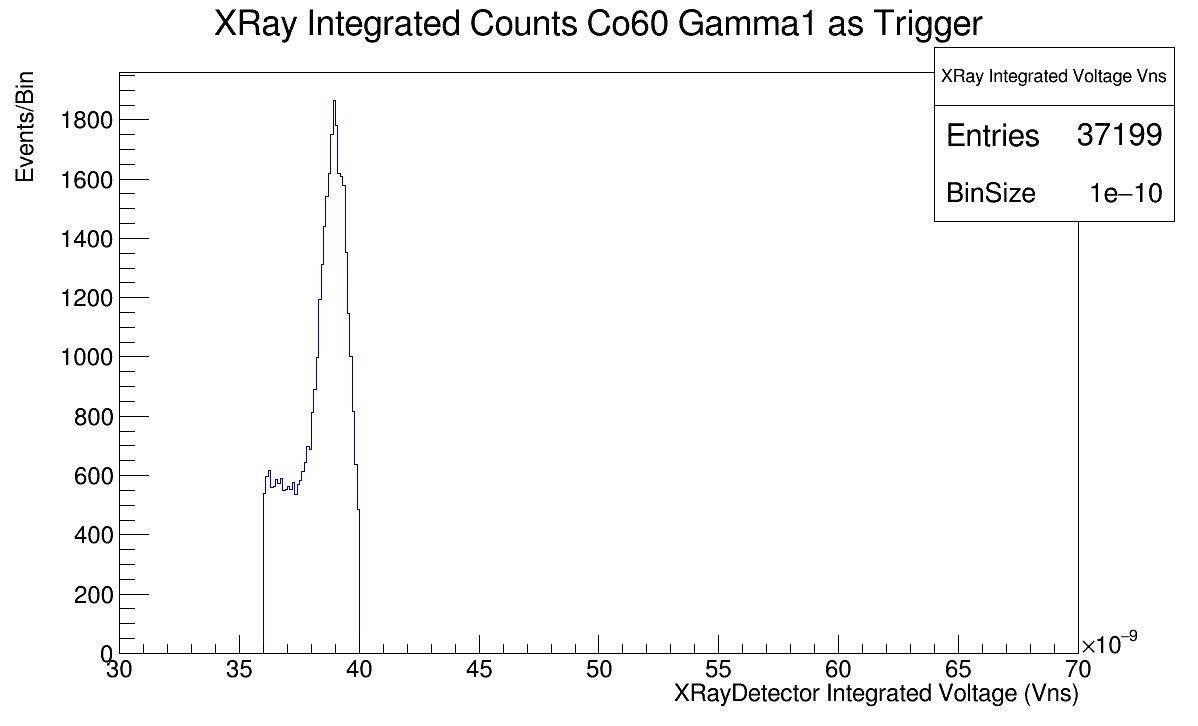}
\caption[Uncut Co60 Data with Trigger1]{Data from $^{60}$Co. (Left) Data restricted to the range corresponding to full energy deposits of gamma 1 in the X-ray detector.}
\label{fig:Co60 Uncut1} 
\end{figure}

\begin{figure}[htbp]
\centering
\includegraphics[width = 0.9\textwidth]{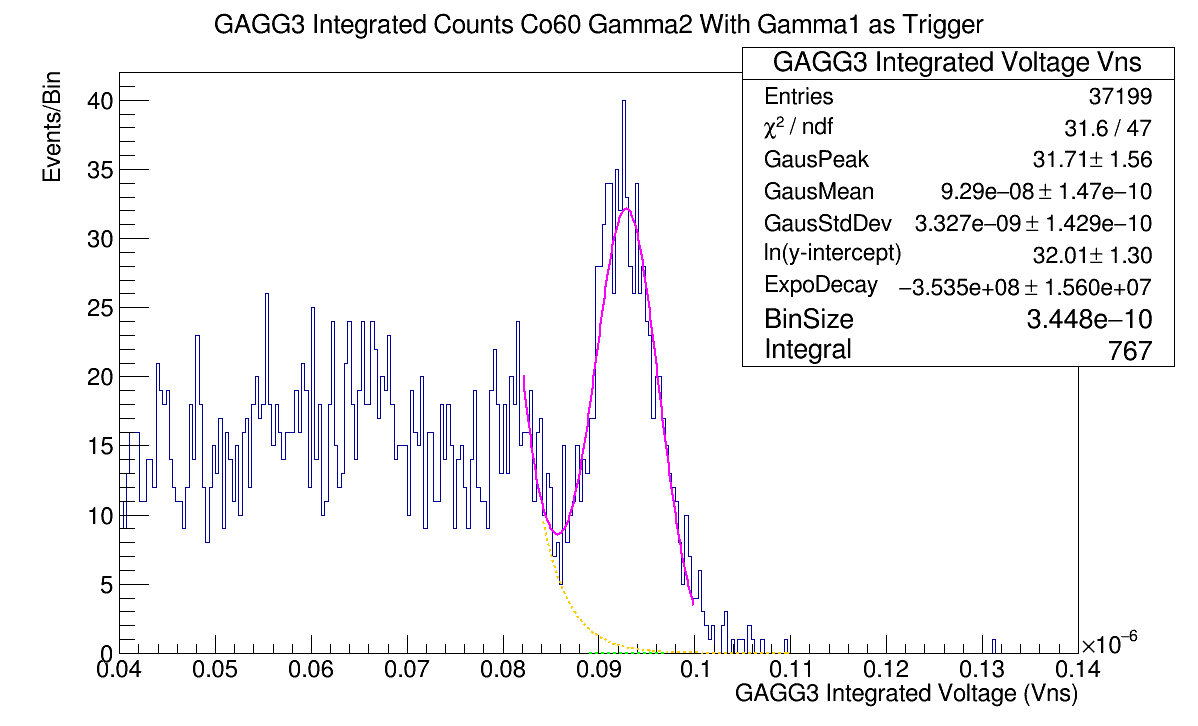}
\caption[Cut Co60 Data with Trigger1]{Data from $^{60}$Co. The events in the GAGG detector when the data is restricted to gamma 1 triggers in the X-ray detector.}
\label{fig:Co60 Cut1} 
\end{figure}

\begin{figure}[htbp]
\centering
\includegraphics[width = 0.9\textwidth]{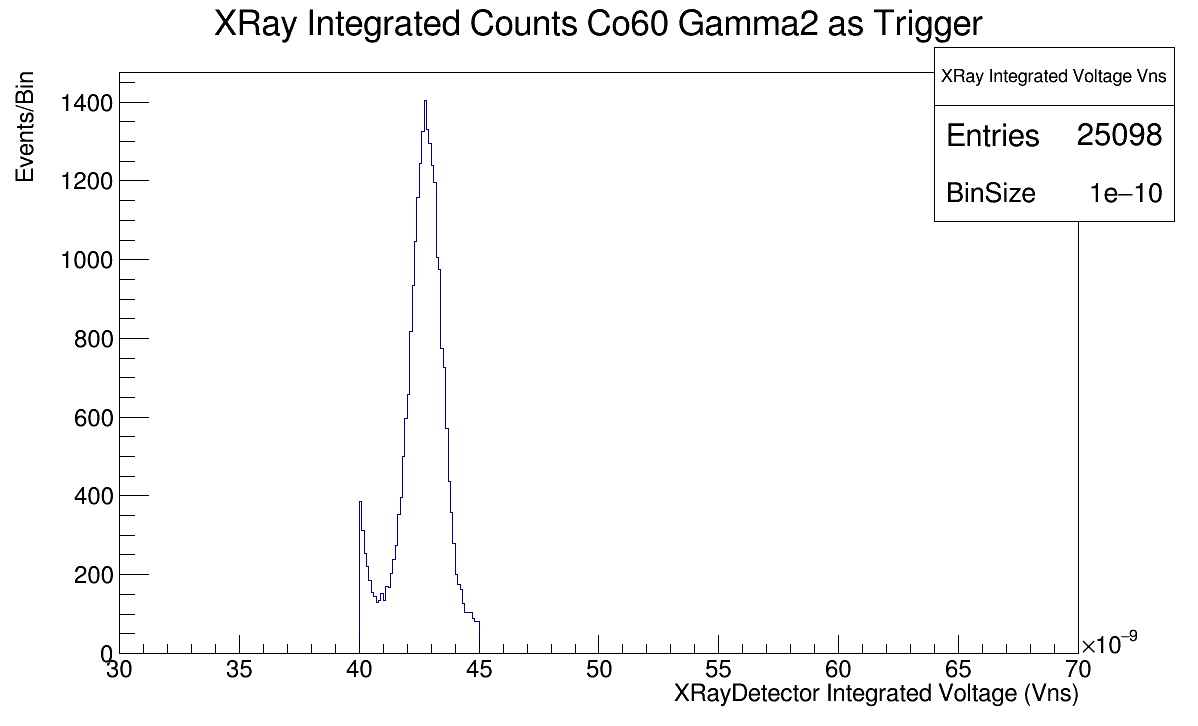}
\caption[Uncut Co60 Data with Trigger2]{Data from $^{60}$Co. Data restricted to the range corresponding to full energy deposits of gamma 2 in the X-ray detector.}
\label{fig:Co60 Uncut2} 
\end{figure}

\begin{figure}[htbp]
\centering
\includegraphics[width = 0.9\textwidth]{ 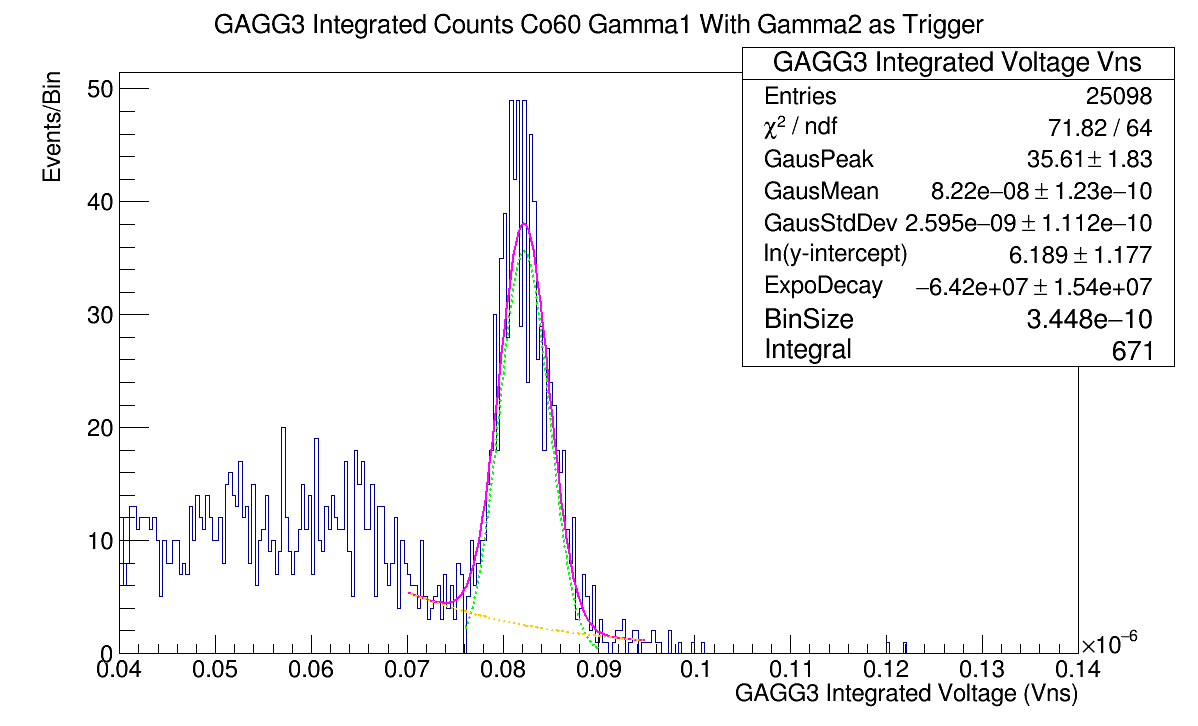}
\caption[Cut Co60 Data with Trigger2]{Data from $^{60}$Co. The events in the GAGG detector when the data is restricted to gamma 2 triggers in the X-ray detector.}
\label{fig:Co60 Cut2} 
\end{figure}

\begin{table}[htbp]
    \centering
        \caption{The final measured vs. simulated efficiency measurement for the flight detector without uncertainties.}
    \begin{tabular}{lll}
        \textbf{Source} & \textbf{Energy} &\textbf{Efficiency} \\\hline\hline
        $^{65}$Zn & 1115 keV & 82.9\% \\\hline
        $^{54}$Mn & 835 keV & 77.4\% \\\hline
        $^{22}$Na & 1274 keV & 90.2\% \\\hline 
        $^{22}$Na & 1785 keV & 90.9\% \\\hline 
        $^{60}$Co & 1173 keV & 91.4\% \\\hline 
        $^{60}$Co & 1332 keV & 82.8\% \\\hline  
        
    \end{tabular}
    \label{tab:FinalEfficiency}
\end{table}

\section{Uncertainties}\label{sec:Uncertainties}

\subsection{Statistical Uncertainties}
\subsubsection{Statistical Uncertainties in measurement}

For both the detected and expected values, we use the uncertainties in the fit parameters to determine the statistical uncertainties by integrating over the parameters with their values adjusted by their 1-$\sigma$ uncertainties. The reason to use 1-$\sigma$ uncertainties is to be consistent with the $\pm$ uncertainties in the fit parameters provided by ROOT with its default confidence value.

In the case of $^{65}$Zn we see a 1-$\sigma$ range of detected X-rays from 133243 to 134738, a fractional uncertainty of -0.57\% and +0.55\%. While the 1-$\sigma$ range in the detected gammas in the flight detector is from 2135 to 2330, a fractional uncertainty of -4.30\% and +4.43\%. 

In the case of $^{54}$Mn, we see a 1-$\sigma$ range of detected X-rays from 53946 to 55135, a fractional uncertainty of -1.19\% and +0.99\%. While the 1-$\sigma$ in the number of 835 keV gamma rays in the flight detector ranges from 2264 to 2472, a fractional uncertainty of -4.35\% and +4.44\%. 

For the $^{22}$Na annihilation gammas in the X-ray detector, the 1-$\sigma$ range is from 111280 to 118783, a fractional uncertainty of -3.18\% and +3.35\%. In the flight detector, the number of detected 1274 keV gammas has a 1-$\sigma$ range from 3885 to 4546, a fractional uncertainty of -7.72\% and +7.98\%. The 1-$\sigma$ range in the combined peak is from 635 to 812, a fractional uncertainty of -11.9\% and +12.6\%.

In the case of $^{60}$Co and specifically for triggering on gamma 1, the 1-$\sigma$ range in the number of detected gammas in the X-ray detector is from 27100 to 28955, a fractional uncertainty of -3.27\% and +3.36\%. For the data set that is restricted to gamma 1 triggers in the X-ray detector, the 1-$\sigma$ range in detected gamma 2 in the flight detector is from 698 to 839, a fractional uncertainty of -9.00\% and +9.39\%.

For the $^{60}$Co triggering on gamma 2 in the X-ray detector, the 1-$\sigma$ range in the number of detected gamma in the X-ray detector is from 19104 to 20007, a fractional uncertainty of -2.28\% and +2.34\%. For the signal in the flight detector when restricted to this data set, the 1-$\sigma$ range in the number of detected gamma 1 is from 610 to 737, a fractional uncertainty of -9.09\% and +9.84\%.

\subsubsection{Statistical Uncertainties in Simulation}

We use the same technique of integrating over the lowest and highest case of fit parameters to obtain the 1-$\sigma$ bounds on the correction factor for each simulation.

As an example: for the $^{65}$Zn 1115 keV gamma ray, the 1-$\sigma$ range in the number of full energy deposits is from 15847 to 16254, a fractional uncertainty of -0.96\% and +1.58\%.

The rest of the values are tabulated in table \ref{tab:Statistical Uncertainty in Correction Factors}






\begin{table}[htbp]
    \centering
        \caption{The statistical uncertainties in the correction factors for each source, all at a 1-$\sigma$ confidence level.}
    \begin{tabular}{lll}
        \textbf{Source} &\textbf{Simulated Range} & \textbf{Fractional Uncertainty}  \\\hline\hline
        $^{65}$Zn & $^{16254}_{15847}$ & $^{+1.58\%}_{-0.96\%}$ \\\hline
        $^{54}$Mn & $^{22669}_{22083}$ & $^{+1.28\%}_{-1.34\%}$ \\\hline
        $^{22}$Na & $^{7172}_{6783}$ & $^{+2.73\%}_{-2.77\%}$ \\\hline 
        $^{22}$Na & $^{2498}_{2276}$ & $^{+4.65\%}_{-4.65\%}$ \\\hline 
        $^{60}$Co & $^{7453}_{7117}$ & $^{+2.35\%}_{-2.27\%}$ \\\hline 
        $^{60}$Co & $^{6446}_{6135}$ & $^{+2.51\%}_{-2.43\%}$ \\\hline  
        
    \end{tabular}
    \label{tab:Statistical Uncertainty in Correction Factors}
\end{table}

\subsection{Systematic Uncertainties}

The main systematic uncertainty for most of the runs performed is going to be uncertainties in the position of the source. Simulations were done with a $^{65}$Zn disk source that placed it in extreme positions relative to the measured position of the source in the lab. Specifically, an offset position with the source moved diagonally 4mm in the x-y plane, and 2 positions dubbed near and far that move the source 1 mm closer and further respectively.

In the lab the flight detector remains fixed, while the X-ray detector is on a stand that remains mostly at a constant height and the entire stand is slid in and out of position, making 1 mm a rather large amount of variation in height, perhaps larger than a 1-$\sigma$ error. The offset simulation changes the correction factor of the $^{65}$Zn source from 4.00\% to 3.88\%, a fractional uncertainty of 2.9\%. The near and far simulations change the correction factor from 4.00\% to 4.22\% and 3.74\% respectively, a fractional uncertainty of +5.5\% and -6.5\% respectively. The uncertainty from position in the actual lab is estimated to be largely due to the small height variations of about 1mm as opposed to lateral position variations of about a few mm, thus the systematic uncertainty is estimated to be about +5.5\% and -6.5\%.

There are known potential systematic effects that have been investigated and confirmed to be a minuscule correction such as angular correlation of the $^{60}$Co gamma cascade, and the dimensions of the source disk. The solid angle of both the logic trigger volume and the simulated flight detector are both very large from the perspective of the source and so the angular correlations in the gamma cascade are integrated over the solid angles to produce no correlation. As for the dimensions of the source, doubling the radius of the cylindrical disk source produced no significant impact on the correction factor in simulation as well as making the source 10 times thicker.

\subsection{Total Uncertainty}

The combined statistical uncertainty, assuming the 3 sources of statistical uncertainty are independent (simulated correction factor, measured triggers in X-ray detector, detected gammas in flight detector), is calculated for each source in the table below [\ref{tab:Total Uncertainty}].

\begin{table}[htbp]
    \centering
        \caption{The fractional total uncertainties for each source, all at a 1-$\sigma$ confidence level. In order: The source, the energy of the gamma, the statistical uncertainty in the correction factor from simulation, the statistical uncertainty in the number of triggers in the X-ray detector, the statistical uncertainty in the number of detected gammas in the flight detector, the estimated systematic uncertainty for each run, and the final combined statistical uncertainty.}
    \begin{tabular}{llllllr}
        \textbf{Source} & \textbf{Energy}&\textbf{CF-$\sigma$} & \textbf{TRIG-$\sigma$} & \textbf{GAGG-$\sigma$} &\textbf{SYST-$\sigma$} & \textbf{TOTAL-$\sigma$}\\\hline\hline
        $^{65}$Zn & 1115 keV & $^{+1.58\%}_{-0.96\%}$ & $^{+0.55\%}_{-0.57\%}$ & $^{+4.43\%}_{-4.30\%}$ & $^{+5.50\%}_{-6.50\%}$ & $^{+12.1\%}_{-12.3\%}$\\\hline
        $^{54}$Mn & 835 keV & $^{+1.28\%}_{-1.34\%}$ & $^{+0.99\%}_{-1.19\%}$ & $^{+4.44\%}_{-4.35\%}$ & $^{+5.50\%}_{-6.50\%}$ & $^{+12.2\%}_{-13.4\%}$\\\hline
        $^{22}$Na & 1274 keV & $^{+2.73\%}_{-2.77\%}$ & $^{+3.35\%}_{-3.18\%}$ & $^{+7.98\%}_{-7.72\%}$ & $^{+5.50\%}_{-6.50\%}$ & $^{+19.5\%}_{-20.2\%}$\\\hline 
        $^{22}$Na & 1785 keV & $^{+4.65\%}_{-4.65\%}$ & $^{+3.35\%}_{-3.18\%}$ & $^{+12.6\%}_{-11.9\%}$ & $^{+5.50\%}_{-6.50\%}$ & $^{+26.1\%}_{-26.2\%}$\\\hline 
        $^{60}$Co & 1173 keV & $^{+2.35\%}_{-2.27\%}$ & $^{+2.34\%}_{-2.28\%}$ & $^{+9.84\%}_{-9.09\%}$ & $^{+5.50\%}_{-6.50\%}$ & $^{+20.0\%}_{-20.1\%}$\\\hline 
        $^{60}$Co & 1332 keV & $^{+2.51\%}_{-2.43\%}$ & $^{+3.36\%}_{-3.27\%}$ & $^{+9.39\%}_{-9.00\%}$ & $^{+5.50\%}_{-6.50\%}$ & $^{+20.8\%}_{-21.2\%}$\\\hline  
        
    \end{tabular}
    \label{tab:Total Uncertainty}
\end{table}

\section{Conclusion}\label{sec:Summary}
The final result for a GAGG flight detector measured vs. simulated efficiency measurement including uncertainties is summarized in Table \ref{tab:Final Results}.
These three tagged gamma ray techniques of X-ray triggered gamma, annihilation triggered gamma, and gamma cascades are a valuable method of determining the measured vs. simulated efficiency and can be used for other scintillating detectors when necessary.

\begin{table}[htbp]
    \centering
        \caption{The final measured vs. simulated efficiency result for the flight detector with total uncertainty.}
    \begin{tabular}{llrr}
        \textbf{Source} & \textbf{Energy} & \textbf{Efficiency} & \textbf{Uncertainty}\\\hline\hline
        $^{65}$Zn & 1115 keV & 82.9\% & $^{+12.1\%}_{-12.3\%}$\\\hline
        $^{54}$Mn & 835 keV  & 77.4\% & $^{+12.2\%}_{-13.4\%}$\\\hline
        $^{22}$Na & 1274 keV & 90.2\% & $^{+19.5\%}_{-20.2\%}$\\\hline
        $^{22}$Na & 1785 keV & 90.9\% & $^{+26.1\%}_{-26.2\%}$\\\hline
        $^{60}$Co & 1173 keV & 91.4\% & $^{+20.0\%}_{-20.1\%}$\\\hline 
        $^{60}$Co & 1332 keV & 92.8\% & $^{+20.8\%}_{-21.2\%}$\\\hline         
    \end{tabular}
    \label{tab:Final Results}
\end{table}


\section{Acknowledgements}\label{sec:Acknowledgements}

Funding: This work was supported by the NASA NIAC Program [grant numbers 80NSSC2K1900, 80NSSC18K0868, 80NSSC19M0971]; MSFC CAN [grant number 80MSFC18M0047]; Wichita State University MURPA; NASA EPSCoR PDG.






\end{document}